\documentclass[preprint,,aps,prb,bibnotes,amsmath,amssymb,floatfix]{revtex4-1}  

\usepackage{graphicx}  
\usepackage{dcolumn}   
\usepackage{bm}        

\begin{document}

\widetext

\title{The Fokker-Planck Equation for Lattice Vibration: Stochastic Dynamics and Thermal Conductivity}
                             
\author{Yi Zeng}
\affiliation{Department of Mechanical Engineering, Auburn University, Auburn, Alabama 36849-5341, U.S.A.}
\author{Jianjun Dong}
\email{Correspondence to: dongjia@auburn.edu}
\homepage{http://www.auburn.edu/cosam/faculty/physics/dong}
\affiliation{Department of Physics, Auburn University, Auburn, Alabama 36849-5311, U.S.A.}

\date{\today}

\begin{abstract}
We propose a Fokker-Planck equation (FPE) theory to describe stochastic fluctuation and relaxation processes of lattice vibration at a wide range of conditions, including those beyond the phonon gas limit. Using the time-dependent, multiple state-variable probability function of a vibration FPE, we first derive  time-correlation functions of lattice heat currents in terms of correlation functions among multiple vibrational modes, and subsequently predict the lattice thermal conductivity based on the Green-Kubo formalism. When the quasi-particle kinetic transport theories are valid, this vibration FPE not only predicts a lattice thermal conductivity that is  identical to the one predicted by the  phonon Boltzmann transport equation, but also provides additional microscopic details on the multiple-mode correlation functions. More importantly, when the kinetic  theories become insufficient  due to the breakdown of the phonon gas approximation, this FPE theory remains valid to study the correlation functions among vibrational modes  in  highly anharmonic lattices with significant mode-mode interactions  and/or in disordered lattices with strongly localized modes. At the limit of weak mode-mode interactions, we can adopt  quantum perturbation theories to derive the drift/diffusion coefficients  based on the lattice anharmonicity data derived from first-principles methods. As temperature elevates to the classical regime, we can perform molecular dynamics simulations to directly compute the drift/diffusion coefficients. Because these coefficients are  defined as ensemble averages at the limit of $\delta t \rightarrow 0$, we can implement massive parallel simulation algorithms to  take full advantage of the paralleled high-performance computing platforms. A better understanding of the temperature-dependent drift/diffusion coefficients up to melting temperatures will provide new insights on microscopic mechanisms that govern the heat conduction through anharmonic and/or  disordered lattices beyond the phonon gas model.


\end{abstract}

\maketitle

\section{\label{sec:intro}Introduction}
	The phonon Boltzmann transport equation (BTE) \cite{peierls1929kinetischen,ziman2001electrons} has gained some renewed interests as the default choice of transport theory to compute  lattice thermal conductivity ($\kappa_{Latt}$) of crystalline solids from first-principles\cite{broido2007intrinsic,ward2009ab,Tang4539,esfarjani2011heat,PhysRevLett.106.045901,fugallo2013ab,tang2014thermal}. The theoretical foundation of the phonon BTE is the so-called phonon gas (PG) model \citep{born1954dynamical,srivastava1990physics,chen2005nanoscale,tritt2005thermal}, which assumes that interactions among vibrational modes are weak enough that the numbers of phonons of each mode follow the single-particle Bose-Einstein distribution at equilibrium. As a kinetic theory, the phonon BTE further assumes that (1) each quasi-particle  phonon  travels at a group velocity $\vec{v}_{g}$,  and (2) the lifetime $\tau$ of every phonon is finite because of the scatterings by lattice anharmonicity, lattice defects/disorder, or other particles.  For electronic insulators, the necessary inputs for a phonon BTE calculation are the harmonic phonon spectra  and the phonon scattering terms, both of which can be numerically calculated using first-principles  methods \cite{RevModPhys.73.515, PhysRevB.60.950,PhysRevB.74.014109,PhysRevLett.75.1819,ward2009ab,tang2009pressure,PhysRevB.77.144112, PhysRevB.84.180301,phonopy,phono3py}.  Multiple implementations of  the phonon BTE methods have been reported in recent years\cite{LI20141747, PhysRevLett.110.265506, CHERNATYNSKIY2015196, CARRETE2017351, PhysRevX.6.041013},  and the calculated  results adopting various theoretical and numerical approximations have been systematically bench-marked among themselves and compared with available experimental data. The overall good agreement between the first-principles computational results and available experimental data for a large amount of crystals at moderate temperatures ($T$) establishes the phonon BTE as a practical and robust computational tool to design advanced technology materials with optimized thermal transport properties.

\begin{figure}[ht]
\includegraphics[scale=0.3]{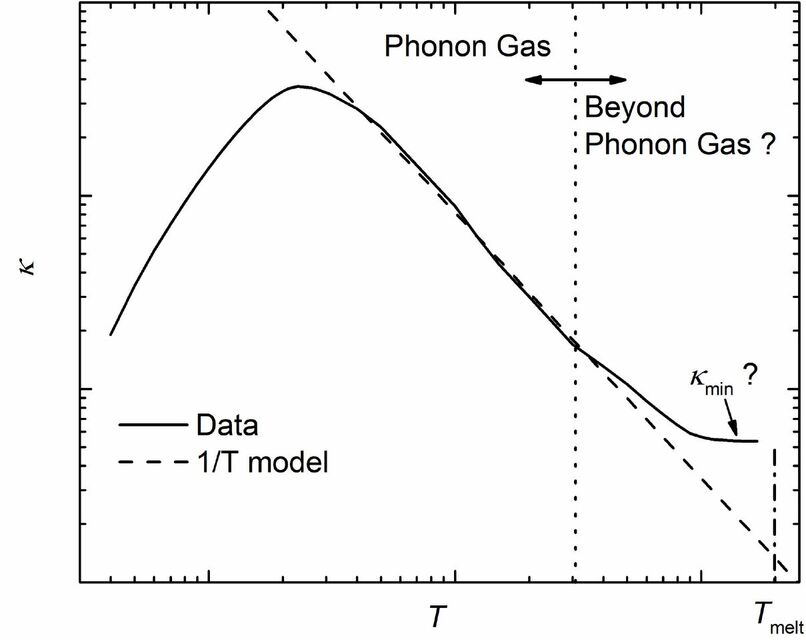}
\caption{Schematic plot of lattice thermal conductivity $\kappa_{Latt}$ as a function of temperature $T$, up to the melting temperature $T_{melt}$.}
\label{fig1}
\end{figure} 

	Meanwhile, concerns have been raised about the validity of the phonon BTE beyond the PG limit, where interactions among vibrational modes are significant and the weakly interacting quasi-particle approximation becomes insufficient \cite{PhysRevB.82.224305}. A schematic plot of a typical temperature dependence of $\kappa_{Latt}$ in crystals is shown in Fig. \ref{fig1}. Within the PG approximation, the phonon BTE predicts that $\kappa_{Latt}$ of a crystal decays to zero with increasing $T$ at the rate of $1/T$ or faster. However, experimental measurements \cite{kappa_si_ge,hofmeister2014thermal}reveal that the deviation from the $1/T$ scaling become noticeable as $T$ approaches the melting temperature ($T_{melt}$) of the lattice, with $\kappa_{Latt}$ eventually reaching a low constant value. The omnipresence of these minimal thermal conductivities ($\kappa_{min}$) \cite{cahill1990thermal} in all crystalline lattices suggests that as a lattice approaches its $T_{melt}$, the increasingly strong  anharmonic coupling among vibrational modes causes the breakdown of the PG model. Such breakdown  might occur at moderate temperatures in relatively soft solids with large thermal expansion \cite{bagieva2012influence,PhysRevB.92.020301,PhysRevB.97.174304}, or in the  high temperature phases of solids whose 0K phonon spectra contain imaginary frequencies \cite{PhysRevX.6.041061}.  In addition, the phonon BTE incorporates  the concept of phonon group velocity, which is not properly defined in non-periodic solids such as alloys, glasses or amorphous semiconductors \cite{ludlam2005universal}, even at the conditions where all the vibrational modes remain quasi-harmonic \cite{allen1993thermal}.
	
	When the accuracy of the phonon BTE theory is in question, the statistical linear response transport theory \cite{kubo2012statistical} is often combined with equilibrium molecular dynamics (MD) simulations to predict thermal transport properties \cite{volz1999molecular,PhysRevLett.86.2361,PhysRevB.79.064301,PhysRevLett.103.015901}. For example, the Green-Kubo (GK) formalism states that  thermal conductivity is proportional to the time-integral of the auto-correlation function of heat flux \cite{green1954markoff, kubo1957statistical}.  Although the GK method is theoretically rigorous and valid beyond the PG approximation, its current implementations, based on the evaluations of atomic trajectories, i.e. displacements and velocities,  over a long period of time,  usually require much more intensive computational loads.  When no reliable empirical force-field interatomic potentials exist, $\it{ab \ initio}$ MD simulations are necessary to simulate the complex lattice vibration. Yet, in practice, typical $\it{ab \ initio}$ MD simulations are often carried out with only relatively short simulation periods (i.e. on the order of a few pico-seconds) and using relatively small super-cell models (i.e. on the order of a couple of hundred atoms) because their computational loads scale as order $N^3$, where $N$ is the number of atoms in a supercell model.  These numerical finite-size artifacts sometimes impose relatively large uncertainties in the  $\it{ab \ initio}$ MD simulation results.  Additional approximations are often needed to extract potential energy of each atom from the $\textit{ab initio}$ total energies of the supercell models in order to evaluate the correlation function of heat currents using the $\textit{ab initio}$ MD simulation results.  \cite{marcolongo2016microscopic,kang2017first,kinaci2012calculation,tse2018thermal}.
	
	More importantly,  all the atomic trajectories in MD simulations have to be calculated numerically, even at the weak scattering limit of the PG model.   This lack of analytical solutions of atomic trajectories in MD simulations hinders the development of  quantitative  theoretical models  to interpret the simulated  current-current  correlation functions because it provides little insights on  improving/correcting the PG model beyond the weak scattering limit.  Ladd  $\textit{et al}$ \cite{PhysRevB.34.5058}  proposed a normal mode analysis (NMA) approach to evaluate  the phonon lifetimes $\tau$  based on the damped oscillator approximation (DOA). Using the extracted phonon lifetimes, they derived the so-called Peierls phonon-transport expression of  $\kappa_{Latt}$, which is understood to be only an approximate solution of  the phonon BTE theory.  Nevertheless, these types of NMA methods have been useful to interpret the phonon scattering in a MD simulation, and these methods have been implemented and  further developed in recent years  by many groups using both empirical potentials and $\it{ab \ initio}$ methods \cite{mcgaughey2004quantitative,henry2008spectral,PhysRevLett.103.125902,dickel2010improved}. However, both the DOA and the concept of phonon lifetime/relaxation-time should be adopted only as semi-quantitative models  because the cross-correlations among different vibrational modes can not always be neglected. More robust theoretical models or concepts are needed to quantitatively interpret the NMA results of numerical MD simulations.  

	In this paper, we present a time-dependent statistical  theory to quantitatively describe the thermal fluctuation and correlation properties of vibrational modes using a Fokker-Planck equation \cite{risken1996fokker} for lattice dynamics. First, this vibration FPE theory does not treat the interactions among different vibrational modes as small perturbations. Instead, our theory includes two general sets of parameters, the drift $A$ and the diffusion  $B$ coefficients, to explicitly characterize the mode-mode interactions. The results of this vibration FPE, expressed in terms of a time-dependent probability function of multiple-variable vibrational micro-states, provide details of the dynamic relaxation processes of lattice vibration, and are readily used by the linear response transport theory to  compute $\kappa_{Latt}$ beyond the quasi-harmonic PG model.
	
	Second, this vibration FPE  provides detailed information on the time-correlation properties of physical quantities without requirement of long time MD simulations. The proposed vibration FPE derives the  correlation functions based on the probability function governed by the drift $A$ and diffusion $B$ coefficients, which are defined in terms of  ensemble averages at the $\delta t \rightarrow 0$ limit. It is important to emphasize that  no \textit{a priori}   forms of correlation  functions are assumed in a FPE calculation of correlation functions. As a result,  when implemented  with first-principles methods, this vibration FPE is promising to be both accurate and  efficient to predict $\kappa_{Latt}$ of novel and complex solids at wide-ranging conditions.
	
	Finally, the $\kappa_{latt}$ predicted by the vibration FPE converges to the one from the conventional phonon BTE within the PG model. Because the FPE's parameters of a lattice vibration can be evaluated with either perturbative methods or simulation methods at the PG approximation, our vibration FPE theory establishes a systematical computational methodology  to analyze errors of the simple PG model and to delineate the breakdown conditions of the PG approximation.

\section{\label{sec:sd}Stochastic Dynamics of Lattice Vibration}
\subsection{\label{sec:fpe}Fokker-Planck equation}	
	The first fundamental assumption of this proposed  Fokker-Planck equation  for lattice vibration is that thermal lattice dynamics is a stochastic process at the microscopic level, and the probabilitic transition dynamics from one vibration micro-state $\Gamma$  to other thermally accessible micro-states  can be modeled with a statistical master equation \cite{kubo2012statistical,risken1996fokker}.  When a specific micro-state $\Gamma^0$ is sampled at time $t =0$,  the initial probability function is simply:
\begin{equation} \label{eq:initial_prob}
P(\Gamma, t=0 \vert \Gamma^0) = \delta(\Gamma - \Gamma^0).
\end{equation}

Regardless of the dynamic details of a stochastic process, the equilibrium ensemble theory constrains that at the long time limit of $t \rightarrow \infty$, the probability function evolves into the canonical distribution function:
\begin{equation} \label{eq:eq_prob}
P(\Gamma, t \rightarrow\infty \vert \Gamma^0) \rightarrow  P_{eq}(\Gamma) = \dfrac{e^\frac{-E(\Gamma)}{k_BT}}{Z_{eq}(T)}, 
\end{equation} where $k_{B}$ is the Boltzmann constant, $T$ represents temperature, $E(\Gamma)$ denotes the energy of any micro-state $\Gamma$, and $Z_{eq}(T)$ denotes the equilibrium canonical partition function of the lattice vibration. The evolution of  this  probability function $P(\Gamma, t \vert \Gamma^0)$ provides a general and quantitative description of lattice thermal relaxation processes,  from a single initially sampled  micro-state $\Gamma^0$ to a set of the thermally accessible micro-states that correspond to an equilibrium distribution governed by the equilibrium statistics. Here, the ergodic condition in lattice vibration is assumed.
	
	We further adopt the Born-von-Karman periodic boundary condition \cite{born1954dynamical} to specify the vibrational micro-states with total $N$ vibration modes, with $N \rightarrow \infty$ for an infinitely large crystal. Using the numbers of  phonons at these modes, i.e. $n_\alpha$ with $\alpha=1,2,3 \cdots,N$,  we specify  a vibrational micro-state with a set of $N$-dimensional state-variables $\Gamma = \{ n_1, n_2, \cdots, n_N \}$. Through the Kramers-Moyal expansion of the master equation,  the time-evolution of this  probability function $P(\Gamma,t|\Gamma^0) = P(n_1, n_2, \cdots, n_N, t \vert n_1^0, n_2^0, \cdots, n_N^0)$ can be expressed in the form of a FPE \cite{kubo2012statistical, risken1996fokker}:
\begin{equation}
\label{eq:phFPE}
\dfrac{\partial P}{\partial t} = - \sum_{\alpha=1}^{N} \dfrac{\partial}{\partial n_\alpha}[A_\alpha(\Gamma)\cdot P] + \dfrac{1}{2} \sum_{\alpha\beta} \dfrac{\partial^2}{\partial n_\alpha \partial n_\beta}[B_{\alpha\beta}(\Gamma) \cdot P].  
\end{equation} The assumption of a FPE is that the third order expansion  coefficients are approximately zero. According to the Pawula theorem, all the higher order expansion coefficients are zero if the third order expansion coefficients are zero \cite{risken1996fokker}. Within this theoretical framework,  the drift $A_{\alpha}(\Gamma)$ and diffusion  $B_{\alpha \beta}(\Gamma)$ coefficients manifest the interactions among vibrational modes, and they are defined as:
\begin{equation} \label{eq:fpe_AB}
\begin{split}
A_{\alpha}(\Gamma) & \equiv \lim_{\delta t \rightarrow 0} \frac{1}{\delta t} \int_{0}^{\delta t} d\Gamma^{'} \delta n_\alpha(\Gamma, \Gamma^{'})P(\Gamma^{'},\delta t \vert \Gamma), \\
B_{\alpha\beta}(\Gamma) & \equiv \lim_{\delta t \rightarrow 0} \frac{1}{\delta t} \int_{0}^{\delta t} d\Gamma^{'} \delta n_\alpha(\Gamma, \Gamma^{'}) \delta n_\beta(\Gamma,\Gamma^{'})P(\Gamma^{'},\delta t \vert \Gamma).\\
\end{split} 
\end{equation}

	Within this statistical probability theory (Eq. \ref{eq:phFPE}), the dynamic details of a stochastic lattice vibration rely on the knowledge of both drift $A$ and diffusion $B$ coefficients. As formulated in Eq. \ref{eq:fpe_AB}, both $A$ and $B$ coefficients can be numerically calculated based on an ensemble of microscopic simulations over a short period of simulation time $\delta t$. Because of the short simulation periods for the parameter evaluation is short, it becomes practical to implement the numerical simulations using accurate first-principles methods. The overall computational loads of ensemble average, although still intensive,  can be in principle distributed over a cluster of computer nodes to take full advantage of the state-of-the-art parallel high-performance computing platforms. Choosing an appropriate  simulation period $\delta t$ for the parameter calculations  is not merely a numeric issue. The length of  $\delta t$ reflects the level of temporal  coarse-graining. For example, in a bulk system, $\delta t$  should be larger than the oscillating periods, as well as  the ballistic time periods, to ensure the assumption of a thermal relaxation process. In addition, different values $\delta t$ might be needed when there are more than one drift/diffusion mechanism. In an amorphous lattice, the drift/diffusion time scale for an extended vibrational mode likely differs significantly from that of a strongly localized vibration mode. Extensive future  studies are needed to gain a better understanding these coefficients of a vibration FPE.
	
	The general forms for the $A$ and $B$ coefficients defined in Eq. \ref{eq:fpe_AB} imply that our proposed vibration FPE theory does not limit the magnitude of the mode-mode interactions in a lattice to be perturbatively small, nor does it require each mode correspond to a traveling wave with a specific group velocity $\vec{v_{\alpha}}$. Consequently, this vibration FPE, as formulated in Eq. \ref{eq:phFPE}, is valid for lattice vibration with a broad range of mode-mode interactions, including  lattice vibration with strong anharmonic modes and/or disorder-induced spatially localized modes. 
	
	Based on the assumption  that a stochastic lattice vibration can be approximated as a random process of transition from one vibrational micro-state $\Gamma$ to another micro-state $\Gamma^{'}$ with a known rate of transition $w_{\Gamma \rightarrow \Gamma{'}}$,  Eq. \ref{eq:fpe_AB} can be approximated as:
\begin{equation} \label{eq:fpe_AB_perturbation}
\begin{split}
A_{\alpha}(\Gamma) & \approx  \int d\Gamma^{'} [n_\alpha(\Gamma^{'}) - n_\alpha(\Gamma)] \cdot w_{\Gamma \rightarrow \Gamma{'}}, \\
B_{\alpha \beta}(\Gamma) & \approx  \int d\Gamma^{'} [n_\alpha(\Gamma^{'}) - n_\alpha(\Gamma)] \cdot  [n_\beta(\Gamma^{'}) - n_\beta(\Gamma)] \cdot w_{\Gamma \rightarrow \Gamma{'}}. \\
\end{split} 
\end{equation}
Within the PG model, both initial ($\Gamma$) and final ($\Gamma^{'}$) quantum vibration states can be represented by the phonon representation $|n_{1}, n_{2}, n_{3}, \cdots, n_{N} >$, and $\Delta \hat{V}$ denotes perturbatively small deviations in the vibration Hamiltonian from that of the ideal phonon gas. We can use Fermi's golden rule to calculate the rate of transition:
\begin{equation} \label{eq:goldenrule}
w_{\Gamma \rightarrow \Gamma{'}} = \frac{2 \pi}{\hbar}|<n^{'}_{1}, n^{'}_{2}, n^{'}_{3}, \cdots, n^{'}_{N}|\Delta \hat{V} | n_{1}, n_{2}, n_{3}, \cdots, n_{N} >|^{2}.
\end{equation}

\subsection{\label{sec:relaxation}Thermal relaxation: fluctuation and correlation}		
	At thermal equilibrium,  the instantaneous value of  a quantity $X$, either macroscopic or microscopic, fluctuates around its equilibrium value $X_{eq}$. The dynamical process that brings the fluctuating  value of $X(t)$ back toward the $X_{eq}$ is commonly referred as  a thermal relaxation process. A self-correlation function of  $X$:
\begin{equation} \label{eq:corr_eq}
C_{XX}(t) \equiv \langle \delta X(0) \cdot \delta X(t) \rangle_{eq} = \langle (X(0) - X_{eq}) \cdot (X(t) - X_{eq}) \rangle_{eq}, 
\end{equation} is often used to  quantify the properties of this thermal relaxation process.  When $X$ can be expressed in terms of micro-state variables $X(\Gamma)$,  we can define a time-dependent expectation value $\overline{X}(t|\Gamma^0)$ based on the probability function $P(\Gamma,t|\Gamma^0) $ in the vibration FPE, staring with the initial probability function shown in Eq. \ref{eq:initial_prob}: 
\begin{equation} \label{eq:average}
\begin{split}
\overline{X}(t|\Gamma^0)  & \equiv \int d\Gamma P(\Gamma,t|\Gamma^0) \cdot X(\Gamma), \\
\frac{d\overline{X}(t|\Gamma^0)}{dt} & \equiv \int d\Gamma \frac{\partial  P(\Gamma,t|\Gamma^0)}{\partial t} \cdot X(\Gamma)\\
 & = \sum_{\alpha} \overline{[\frac{\partial X}{\partial n_{\alpha}} \cdot A_\alpha ]} (t|\Gamma^0) + \dfrac{1}{2}\sum_{\alpha \beta} \overline{[\frac{\partial^2 X}{\partial n_{\alpha} \partial n_{\beta}} \cdot B_{\alpha \beta}]} (t|\Gamma^0).
\end{split}
\end{equation} Clearly, $\overline{X}(t)$ starts at its initial value of $X(\Gamma^0) = \int d\Gamma \delta(\Gamma-\Gamma^0) X(\Gamma)$,  and eventually relaxes back to its equilibrium value of $X_{eq} = \int d\Gamma X(\Gamma) P_{eq}(\Gamma)$ when $P(\Gamma,t | \Gamma^0) \rightarrow P_{eq}(\Gamma)$ at the limit of $t \rightarrow \infty$. Similarly, the  corresponding time-dependent statistical variance, defined as  $\Delta_{X}(t|\Gamma_0)  \equiv \overline{X^2}(t|\Gamma_0) - \overline{X}(t|\Gamma_0)^2$,  relaxes from its initial value of 0 to its equilibrium value  $\Delta_{X,eq}=\int d\Gamma (X(\Gamma) - X_{eq})^2 \cdot P_{eq}(\Gamma) >0$.

	By sampling the initial micro-states $\Gamma^0$ with the equilibrium probability function $P_{eq}(\Gamma^0)$,  we can re-write the time-correlation function of $X$,   defined in Eq. \ref{eq:corr_eq}, as:
\begin{equation} \label{eq:correlation}
\begin{split}
C_{XX}(t) & = \langle \delta X(\Gamma_0) \cdot \overline{\delta X}(t|\Gamma_0) \rangle_{eq} \\
& = \int d\Gamma_0  P_{eq}(\Gamma_0)(X(\Gamma_0)-X_{eq}) \int d\Gamma  P(\Gamma,t|\Gamma_0) (X(\Gamma)-X_{eq}), 
\end{split}
\end{equation} where $C_{XX}(t=0)=\Delta_{X,eq}$, and $C_{XX}(t \rightarrow \infty)  \rightarrow  [\int d\Gamma_0 (X(\Gamma_0)-X_{eq}) \cdot P_{eq}(\Gamma_0)] \cdot [ \int d\Gamma  P_{eq}(\Gamma)  \cdot (X(\Gamma)-X_{eq})] = 0$. A concept of an effective relaxation time ($\tau_{X}$) of $X$ is frequently adopted as the time integration of the normalized self-correlation function $c_{XX}(t) \equiv C_{XX}(t)/\Delta_{X,eq}$:
\begin{equation} \label{eq:effective_tau}
\tau_{X} \equiv \int_{0}^{\infty}  c_{XX}(t) dt,
\end{equation} based on the approximation that $c_{XX}(t) \approx e^{-t/\tau_X}$. 

	The dynamical  correlation  between two different quantities $X$ and $Y$ that fluctuate around their prospective equilibrium values ($X_{eq}$ and $Y_{eq}$ can be quantitatively formulated in terms of  a cross-correlation function $C_{XY}(t)$:
\begin{equation} \label{eq:cross_corr_eq}
C_{XY}(t) \equiv \langle \delta X(0) \cdot \delta Y(t) \rangle_{eq} =  \langle (X(0) - X_{eq}) \cdot (Y(t) - Y_{eq}) \rangle_{eq}, 
\end{equation} and this cross-correlation function can be re-written using the probability distribution function  $P(\Gamma,t|\Gamma_0)$ of Eq. \ref{eq:phFPE} :
\begin{equation} \label{eq:correlation_cross}
\begin{split}
C_{XY}(t) & = \langle \delta X(\Gamma_0) \cdot \overline{\delta Y}(t|\Gamma_0) \rangle_{eq} \\
& = \int d\Gamma_0  P_{eq}(\Gamma_0) (X(\Gamma_0)-X_{eq}) \int d\Gamma P(\Gamma,t|\Gamma_0) (Y(\Gamma)-Y_{eq}), 
\end{split}
\end{equation} where $C_{XY}(t \rightarrow \infty)  \rightarrow  [\int d\Gamma_0 (X(\Gamma_0)-X_{eq}) \cdot P_{eq}(\Gamma_0)] \cdot [ \int d\Gamma  P_{eq}(\Gamma)  \cdot (Y(\Gamma)-Y_{eq})] = 0$.   Since  $C_{XY}(t=0)= \int d\Gamma_0  P_{eq}(\Gamma_0) (X(\Gamma_0)-X_{eq}) \cdot (Y(\Gamma_o)-Y_{eq}) = \langle (X - X_{eq}) \cdot (Y - Y_{eq}) \rangle_{eq} $,  the ratio   $c_{XY}  \equiv  C_{XY}(t=0)/{\sqrt{\Delta_{X,eq} \cdot \Delta_{Y,eq}}}$ is often referred as the correlation ratio, with $c_{XY}=0$ being interpreted as that the fluctuations in $X$ and $Y$ are statistically uncorrelated at thermal equilibrium. It is important to emphasize that even at the condition of zero correlation ration, i.e.   $c_{XY}=0$, a cross-correlation function defined in Eq. \ref{eq:correlation_cross} in not always zero at $t > 0$.

	Because the self-correlation function formula in Eq. \ref{eq:correlation} is a special case of the cross-correlation function formula  in Eq. \ref{eq:correlation_cross} with $X=Y$,  we present only the results of the time derivative of the cross-correlation function here based on  Eqs. \ref{eq:average} and \ref{eq:correlation_cross}:
\begin{equation} \label{eq:dt_dCxy}
\begin{split}
\frac{dC_{XY}(t)}{dt} & =  \int d\Gamma_0  P_{eq}(\Gamma_0) \delta X(\Gamma_0) \frac{d \overline{\delta Y}(t|\Gamma_0)}{dt} \\
& = \int d\Gamma_0  P_{eq}(\Gamma_0) \delta X(\Gamma_0) \cdot \{\sum_{\mu} \overline{[\frac{\partial Y}{\partial n_{\mu}} \cdot A_\mu ]} (t|\Gamma^0) + \dfrac{1}{2}\sum_{\mu \nu} \overline{[\frac{\partial^2 Y}{\partial n_{\mu} \partial n_{\nu}} \cdot B_{\mu \nu}]} (t|\Gamma^0) \}. \\
\end{split}
\end{equation} where $A$ and $B$ are the parameters (Eq. \ref{eq:fpe_AB} ) of the vibration FPE (Eq. \ref{eq:phFPE}). Using the definitions of $y_{\mu} \equiv \frac{\partial Y}{\partial n_{\mu}} \cdot A_\mu$ and $y_{\mu \nu} \equiv \frac{\partial^2 Y}{\partial n_{\mu} \partial n_{\nu}} \cdot B_{\mu \nu}$, we can re-write Eq. \ref{eq:dt_dCxy} in terms of the cross-correlation functions between $X$ and $y_{\mu}$ and those between $X$ and $y_{\mu \nu}$:
\begin{equation} \label{eq:dt_dCxy_2}
\begin{split}
\frac{dC_{XY}(t)}{dt}  & = \sum_{\mu} \langle \delta X(0) \cdot y_{\mu}(t) \rangle _{eq} + \dfrac{1}{2}\sum_{\mu \nu} \langle \delta X(0) \cdot y_{\mu \nu}(t) \rangle _{eq} \\
& = \sum_{\mu} \langle \delta X(0) \cdot \delta y_{\mu}(t) \rangle _{eq} + \dfrac{1}{2}\sum_{\mu \nu} \langle \delta X(0) \cdot \delta y_{\mu \nu}(t) \rangle _{eq} \\
& = \sum_{\mu} C_{X y_{\mu}}(t) + \dfrac{1}{2}\sum_{\mu \nu} C_{X y_{\mu \nu}}(t).\\
\end{split}
\end{equation} Furthermore, all the higher order time derivatives of $C_{XY}(t)$ functions can also be derived from Eq. \ref{eq:dt_dCxy_2} in a recursive fashion.  


	Next, we summarize some key results in the case that $X$ and $Y$ are simply the $\alpha$-th and $\beta$-th state variables $n_\alpha$ and $n_\beta$, with more details  on the mathematical derivation given in Appendix \ref{sec:Aa}. The commonly adopted concept of phonon occupation number of a vibrational mode can be generalized as the time-dependent expectation value of the state variable $n_\alpha$ during a thermal relaxation process, i.e. $\overline{n}_\alpha(t|\Gamma^0)  \equiv \int d\Gamma n_\alpha P(\Gamma, t \vert \Gamma^{0})$, with $\overline{n}_{\alpha}(t|\Gamma^0)  \rightarrow n_{\alpha,eq}$ and $\Delta_{\alpha}(t|\Gamma^0) \equiv \overline{n_\alpha^2}(t|\Gamma^0) - \overline{n_\alpha}(t|\Gamma^0)^2 \rightarrow \Delta_{\alpha,eq} $ at the $t \rightarrow \infty$ limit.  At the weak phonon scattering limit of the PG model, the thermal equilibrium values of $n_{\alpha,eq}$ follow the Bose-Einstein distribution, and the corresponding statistical variances are $\Delta_{\alpha,eq}=n_{\alpha,eq}(n_{\alpha,eq}+1)$. Applying the vibration FPE ( Eq. \ref{eq:phFPE}) to Eq. \ref{eq:average}, we derive the time derivatives of $\overline{n}_\alpha(t|\Gamma^0) $ and $\Delta_{\alpha}(t|\Gamma^0)$ as:
\begin{equation}
\label{eq:dndt}
\begin{split}
\dfrac{d}{dt} \overline{n_{\alpha}}(t|\Gamma^0)  & =  \int d \Gamma A_\alpha(\Gamma)P(\Gamma, t \vert \Gamma^0) =  \overline{A_\alpha}(t|\Gamma^0), \\
\dfrac{d}{dt}\Delta_{\alpha} (t|\Gamma^0)  & = \overline{B_{\alpha \alpha}} (t|\Gamma^0) + 2 \cdot [ \overline{n_\alpha A_\alpha}(t|\Gamma^0)  - \overline{n_\alpha}(t|\Gamma^0) \cdot \overline{A_\alpha}(t|\Gamma^0) ].
\end{split}
\end{equation}
	
	Furthermore, using  Eqs. \ref{eq:cross_corr_eq} and \ref{eq:correlation_cross},  we define the  cross-correlation functions between the fluctuating phonon numbers of the $\alpha$-th mode and the $\beta$-th mode (also referred to as two-mode correlation functions) as $C_{n_{\alpha} n_{\beta}}(t) \equiv \langle \delta n_{\alpha}(0) \cdot \delta n_{\beta}(t) \rangle _{eq} = \langle n_{\alpha}(0) \cdot n_{\beta}(t) \rangle _{eq} - n_{\alpha,eq}  \cdot n_{\beta , eq}$, with $C_{n_{\alpha} n_{\beta}}(t=0) = \langle \delta n_\alpha \cdot \delta n_\beta \rangle _{eq} = \langle n_\alpha \cdot n_\beta \rangle _{eq} -  n_{\alpha,eq}  \cdot n_{\beta , eq}$. We can further define the normalized two-mode correlation functions as:
\begin{equation} \label{eq:normal_twomode}
c_{\alpha \beta}(t) \equiv \frac{C_{n_{\alpha} n_{\beta}}(t)}{\sqrt{\Delta_{\alpha, eq} \cdot  \Delta_{\beta, eq} }} = \frac{\int d\Gamma^0  P_{eq}(\Gamma^0) \delta n_{\alpha}(\Gamma^0) \cdot \overline{\delta n_{\beta}} (t|\Gamma^0)}{\sqrt{\Delta_{\alpha, eq} \cdot  \Delta_{\beta, eq} }}.
\end{equation} Since $X=n_{\alpha}$ and $Y=n_{\beta}$, we have $y_{\mu} = A_{\beta} \cdot \delta_{\mu \beta}$ and $y_{\mu \nu} = 0$. Using Eq. \ref{eq:dt_dCxy_2}, we can show that:
\begin{equation} \label{eq:dt_normal_twomode}
\frac{dc_{\alpha \beta}(t)}{dt} = \dfrac{ C_{n_\alpha A_{\beta}}(t)}{\sqrt{\Delta_{\alpha, eq} \cdot  \Delta_{\beta, eq} }} =  \dfrac{ \langle \delta n_{\alpha}(0) A_{\beta}(t) \rangle _{eq}}{\sqrt{\Delta_{\alpha, eq} \cdot  \Delta_{\beta, eq} }} = \frac{\int d\Gamma^0  P_{eq}(\Gamma^0) \delta n_{\alpha}(\Gamma^0) \cdot \overline{A_{\beta}} (t|\Gamma^0)}{\sqrt{\Delta_{\alpha, eq} \cdot  \Delta_{\beta, eq} }} .
\end{equation}	  
	
	Multiple-mode correlation functions can be defined in a similar fashion. For example, there is only one type of three-mode correlation function among the $\alpha$-th, $\beta$-th, and $\gamma$-th mode:
\begin{equation} \label{eq:three_mode_corr}
\begin{split}
& \langle \delta n_{\alpha}(0) \cdot \delta n_{\beta}(0) \cdot \delta n_{\gamma}(t) \rangle _{eq} =  \int d\Gamma^0  P_{eq}(\Gamma^0) \delta n_{\alpha}(\Gamma^0) \cdot \delta n_{\beta}(\Gamma^0) \cdot \overline{\delta n_{\gamma}} (t|\Gamma^0) =\\
& \int d\Gamma^0  P_{eq}(\Gamma^0)(n_\alpha(\Gamma^0)-n_{\alpha,eq}) \cdot (n_\beta(\Gamma^0)-n_{\beta,eq}) \int d\Gamma P(\Gamma,t|\Gamma^0)  (n_\gamma(\Gamma)-n_{\gamma,eq}),\\
\end{split}
\end{equation}  and there are three types of four-mode correlation functions among four ($\alpha$, $\beta$, $\mu$, and $\nu$) modes:

\begin{equation} \label{eq:four_mode_corr_1}
\begin{split}
& \langle \delta n_\alpha(0) \delta n_\beta(0) \delta n_\mu(0) \delta n_\nu(t) \rangle _{eq} = \int d\Gamma^0  P_{eq}(\Gamma^0) \delta n_{\alpha}(\Gamma^0) \cdot \delta n_{\beta}(\Gamma^0) \cdot \delta n_{\mu}(\Gamma^0) \cdot   \overline{\delta n_{\nu}} (t|\Gamma^0) =\\ 
& \int d\Gamma^0   P_{eq}(\Gamma^0)(n_\alpha(\Gamma^0)-n_{\alpha,eq}) \cdot (n_\beta(\Gamma^0)-n_{\beta,eq}) \cdot (n_\mu(\Gamma^0)-n_{\mu,eq}) \int d\Gamma P(\Gamma,t|\Gamma^0)   (n_\nu(\Gamma)-n_{\nu,eq}),\\
\end{split}
\end{equation}

\begin{equation} \label{eq:four_mode_corr_2}
\begin{split}
& \langle \delta n_\alpha(0) \delta n_\beta(0) \delta n_\mu(t) \delta n_\nu(t) \rangle _{eq} =  \int d\Gamma^0  P_{eq}(\Gamma^0) \delta n_{\alpha}(\Gamma^0) \cdot \delta n_{\beta}(\Gamma^0) \cdot \overline{\delta n_{\mu} \cdot \delta n_{\nu}} (t|\Gamma^0) =\\
&  \int d\Gamma^0  P_{eq}(\Gamma^0) (n_\alpha(\Gamma^0)-n_{\alpha,eq}) \cdot (n_\beta(\Gamma^0)-n_{\beta,eq}) \int d\Gamma P(\Gamma,t|\Gamma^0)   (n_\mu(\Gamma)-n_{\mu,eq}) \cdot (n_\nu(\Gamma)-n_{\nu,eq}),\\
\end{split}
\end{equation}

\begin{equation} \label{eq:four_mode_corr_3}
\begin{split}
& \langle \delta n_\alpha(0) \delta n_\beta(t) \delta n_\mu(t) \delta n_\nu(t) \rangle _{eq} = \int d\Gamma^0  P_{eq}(\Gamma^0) \delta n_{\alpha}(\Gamma^0) \cdot \overline{\delta n_{\beta} \cdot \delta n_{\mu} \cdot \delta n_{\nu}} (t|\Gamma^0) = \\
&  \int d\Gamma^0  P_{eq}(\Gamma^0)(n_\alpha(\Gamma^0)-n_{\alpha,eq}) \int d\Gamma P(\Gamma,t|\Gamma^0)  (n_\beta(\Gamma)-n_{\beta,eq}) \cdot (n_\mu(\Gamma)-n_{\mu,eq}) \cdot (n_\nu(\Gamma)-n_{\nu,eq}).\\
\end{split}
\end{equation}

	Within the PG model, the fluctuations of phonon occupation numbers at two different  modes are considered to be  statistically independent at a thermal equilibrium, i.e. $<n_\alpha \cdot n_\beta>_{eq} = n_{\alpha, eq} \cdot  n_{\beta, eq}$ for $\alpha \neq \beta$. As a result, the values of the normalized time-correlation function at  $t = 0$ are simply $c_{\alpha \beta}(t =0) = \delta_{\alpha \beta}$, where $\delta_{\alpha \beta}$ is the Kronecker-$\delta$ symbol. Yet, the PG model does not state the value of a cross-correlation  function (Eq. \ref{eq:normal_twomode}) at any other time $ t \neq 0$, except that $c_{\alpha \beta}(t) \rightarrow 0$ as $ t \rightarrow \infty$.  Multiple-mode correlation functions remain poorly understood, even within the PG model.



\subsection{\label{sec:oua}Ornstein-Uhlenbeck Processes}
	The FPE for a well-studied class of stochastic processes, the so-called Ornstein-Uhlenbeck (OU) processes \cite{uhlenbeck1930ge}, can be solved analytically. To demonstrate the properties of these OU processes, we start with a new set of zero-mean and unit-variance stochastic variables $\tilde{\Gamma}=(x_1,x_2,x_3, \cdots ,x_N)$, i.e.  $\langle x_{\lambda} \rangle_{eq} = 0$ and $\langle {x_{\lambda}}^2 \rangle_{eq} = 1$. The OU processes are defined in terms of their specific form of drift and diffusion coefficients: $A_\lambda(\tilde{\Gamma})=- \gamma_\lambda x_\lambda$  and  $B_{\lambda \lambda^{'}} (\tilde{\Gamma})=2\gamma_\lambda \delta_{\lambda,\lambda^{'}}$, with $\gamma_\lambda >0$. Consequently, the Fokker-Planck equation for an OU type processes can be re-written in a separable multiple-variable partial differential equation:
\begin{equation}
\label{eq:ou_fpe}
\dfrac{\partial P(\tilde{\Gamma},t |{\tilde{\Gamma}}^0)}{\partial t} =   \sum_{\lambda=1}^{N} \gamma_\lambda [ 1 +  x_\lambda \cdot \dfrac{\partial}{\partial x_\lambda} +  \dfrac{\partial^2}{\partial x_\lambda ^2} ] P(\tilde{\Gamma},t |{\tilde{\Gamma}^0}),    
\end{equation} and its solution can be expresses as: 
\begin{equation} \label{eq:ou_prob}
P(\tilde{\Gamma}, t |{\tilde{\Gamma}^0}) = \prod_{\lambda  =1} ^N \dfrac{1}{\sqrt{2\pi\Delta_\lambda (t)}} e^{-\dfrac{[x_\lambda - \overline{x_{\lambda}}(t)]^2}{2\Delta_\lambda (t)}}, 
\end{equation}
where, $\overline{x_{\lambda}}(t) = x_{\lambda}(\tilde{\Gamma}^0) \cdot e^{-\gamma_\lambda t}$. 
and $\Delta_\lambda (t) = 1 - e^{-2 \gamma_\lambda t}$. More details on the solution of an OU type FPE can be found in Appendix \ref{sec:Ab}. Here we highlight one key result of the time-correlation between any two state variables $x_\lambda$ and $x_{\lambda '}$  of an OU type process: 
\begin{equation} \label{eq:ou_corr}
\tilde{C}_{\lambda \lambda '}(t) = \langle x_{\lambda}(t^{'}) \cdot x_{\lambda '}(t^{'} + t) \rangle _{eq} =  \delta_{\lambda,\lambda^{'}} e^{-\gamma_\lambda t}.
\end{equation} More interesting results on the multiple variable correlation functions, such as the three-variable correlation functions:  $\langle x_{\lambda}(t^{'}) \cdot x_{\lambda '}(t^{'})  \cdot x_{\lambda ''}(t^{'} + t) \rangle _{eq}$, $\langle x_{\lambda}(t^{'}) \cdot x_{\lambda '}(t^{'}+t)  \cdot x_{\lambda ''}(t^{'} + t) \rangle _{eq}$,  and the four-variable correlation functions $\langle x_{\lambda}(t^{'}) \cdot x_{\lambda '}(t^{'})  \cdot x_{\lambda ''}(t^{'})  \cdot  x_{\lambda '''}(t^{'} + t) \rangle _{eq}$ and    $\langle x_{\lambda}(t^{'}) \cdot x_{\lambda '}(t^{'})  \cdot x_{\lambda ''}(t^{'} + t)  \cdot  x_{\lambda '''}(t^{'} + t) \rangle _{eq}$,  $\langle x_{\lambda}(t^{'}) \cdot x_{\lambda '}(t^{'}+t)  \cdot x_{\lambda ''}(t^{'} + t)  \cdot  x_{\lambda '''}(t^{'} + t) \rangle _{eq}$, are presented in Appendix \ref{sec:Ab}. 


	
	
	For a lattice vibration to be classified as an OU process, its set of drift coefficients $A(\Gamma)$ in the vibration FPE (Eq. \ref{eq:phFPE}) must satisfy the following conditions:
\begin{equation} \label{eq:ou_A}
\begin{split}
A_\alpha(\Gamma) = - \sum_{\beta} \mathcal{D}_{\alpha \beta} (\frac{\Delta_{\alpha,eq}}{\Delta_{\beta,eq}})^{1/2} (n_\beta - n_{\beta, eq}), \\
\dfrac{d \overline{n_{\alpha}}(t|\Gamma^0)}{d t} = -  \sum_{\beta} \mathcal{D}_{\alpha \beta} (\frac{\Delta_{\alpha,eq}}{\Delta_{\beta,eq}})^{1/2} (\overline{n_{\beta}}(t|\Gamma^0)) - n_{\beta,eq}) .
\end{split}
\end{equation} Here $\mathcal{D}_{\alpha \beta}$ are matrix elements  of  the normalized drift matrix $\pmb{\mathcal{D}}$, $n_{\alpha, eq}$ and $\Delta_{\alpha,eq}$ are respectively the equilibrium average value of the phonon number at $\alpha$-th mode and the corresponding statistical variance at the equilibrium with  $\alpha, \beta = 1,2,3, \cdots, N$.

	The  $\pmb{\mathcal{D}}$ matrix, as defined in Eq. \ref{eq:ou_A}, is a positive definite, real, and symmetric $N \times N$ matrix with a set of $N$ eigenvalues $\gamma_\lambda$ and corresponding normalized eigenvectors written as as $\vec{u}_\lambda = (u_{\lambda,1}, u_{\lambda,2}, u_{\lambda,3}, \cdots, u_{\lambda,N}) $ for $\lambda = 1,2, 3, \cdots, N$.
We then can transform the $N$-dimensional phonon number state variables $\Gamma = \{ n_1, n_2, \cdots, n_N \}$ into  an equivalent set of zero-mean and unit-variance state variables $\tilde{\Gamma}=(x_1,x_2,x_3, \cdots ,x_N)$ using this set of eigenvectors:
\begin{equation} \label{eq:transform}
\begin{split}
n_\alpha & = n_{\alpha,eq} + (\Delta_{\alpha, eq})^{1/2} \sum_{\lambda =1}^{N} x_\lambda u_{\lambda,\alpha}, \\
x_\lambda & = \sum_{\alpha = 1}^{N} \frac{n_\alpha - n_{\alpha, eq}}{{\Delta_{\alpha, eq}}^{1/2}}u_{\lambda,eq}.
\end{split}
\end{equation}

	The linear transformation in Eq. \ref{eq:transform} also shows that the diffusion  $B_{\alpha \beta}(\Gamma)$ coefficients for an OU type lattice vibration  are related to its drift coefficients $A_\alpha(\Gamma)$ through the $\pmb{\mathcal{D}}$ matrix:
\begin{equation} \label{eq:ou_B}
B_{\alpha \beta}(\Gamma) = 2(\Delta_{\alpha,eq} \cdot \Delta_{\beta,eq})^{1/2} \sum_{\lambda = 1}^{N} \gamma_{\lambda}u_{\lambda,\alpha}u_{\lambda,\beta} = 2(\Delta_{\alpha,eq} \cdot \Delta_{\beta,eq})^{1/2}  \mathcal{D}_{\alpha \beta}. 	
\end{equation} In the rest of the paper,  the $\pmb{\mathcal{D}}$ matrix is referred as the normalized drift/diffusion matrix. 

	Combining the results in Eqs \ref{eq:normal_twomode}, \ref{eq:ou_corr} and \ref{eq:transform}, we can show that the normalized two-mode correlation functions $c_{\alpha \beta}(t)$ (Eq. \ref{eq:normal_twomode})  in this OU type lattice vibration are simply:
\begin{equation} \label{eq:ou_mode_corr}
c_{\alpha \beta}(t) = \sum_{\lambda = 1}^{N} e^{-\gamma_{\lambda}t} u_{\lambda,\alpha} u_{\lambda, \beta}, 
\end{equation} with $c_{\alpha \beta}(t=0) = \sum_{\lambda = 1}^{N} u_{\lambda,\alpha} u_{\lambda, \beta} = \delta_{\alpha \beta}$. We can generalize the normalized two-mode correlation functions in Eq. \ref{eq:ou_mode_corr} in an integral form:
\begin{equation} \label{eq:integral_mode_corr}
c_{\alpha \beta}(t)  = \int_{0}^{\infty} d\gamma \chi_{\alpha \beta}(\gamma)  e^{-\gamma t},
\end{equation}
with $\chi_{\alpha \beta}(\gamma) =  \sum_{\lambda = 1}^{N} u_{\lambda,\alpha} u_{\lambda, \beta} \cdot \delta(\gamma - \gamma_{\lambda})$. Eq. \ref{eq:integral_mode_corr} indicates that a mode correlation function $c_{\alpha \beta}(t)$ can be viewed as the $t$-space Laplace transformation of the $\gamma$-space function $\chi_{\alpha \beta}(\gamma)$.  We refer to $\chi_{\alpha \beta}(\gamma)$ as the Laplace spectral function of $c_{\alpha \beta}(t)$. At the $N \rightarrow \infty$ limit, a Laplace spectral function  $\chi_{\alpha \beta}(\gamma)$ converges to  a continuous function defined in the spectral regime of $[0,\gamma_{max}]$. The $k$-th moment of a $\chi_{\alpha \beta}(\gamma)$ function, defined as $\mu_{\alpha \beta}(k) \equiv \int_{0}^{\infty} d\gamma \chi_{\alpha \beta}(\gamma) \cdot \gamma^{k}$, is given as:
\begin{equation} \label{eq:moments}
\mu_{\alpha \beta}(k) = \sum_{\lambda = 1}^{N} u_{\lambda,\alpha} u_{\lambda, \beta}  {\gamma_{\lambda}}^{k}=<\alpha|\pmb{\mathcal{D}}^{k}|\beta> =(\pmb{\mathcal{D}}^{k})_{\alpha \beta}.
\end{equation}
	
	The results in Eqs. \ref{eq:ou_mode_corr} and \ref{eq:integral_mode_corr} clearly demonstrate that in general the normalized mode self-correlaltion functions of lattice vibration  do not decay as an exponetial function of time, and the time-integral of the cross-correlaltion functions are not zero for two different modes. Some recent simulation studies \cite{PhysRevLett.103.125902} have reported their implementation based on fitting the  MD simulated mode self-correlation  functions based on an assumed formula of  $ C_{\alpha \beta}(t) \approx \Delta_{\alpha,eq} \cdot \delta_{\alpha, \beta} \cdot e^{-\gamma_{\alpha} t}$, and they reported the fitted decay factors $\gamma_{\alpha}$ as the inverse of phonon life-times $\tau_{\alpha}= \gamma_{\alpha}^{-1}$ in the PG model. For such a simplification to be valid, the normalized drift/diffusion matrix $\pmb{\mathcal{D}}$ has to be close to a diagonal matrix:
\begin{equation} \label{eq:diagonal_L}
\quad \  \pmb{\mathcal{D}} \approx
\begin{bmatrix}
    \gamma_{1} & 0 & 0 & \dots  & 0 \\
    0 & \gamma_{2} & 0 & \dots  & 0 \\
    \vdots & \vdots & \vdots & \ddots & \vdots \\
    0 & 0 & 0 & \dots  & \gamma_{N}
\end{bmatrix}
\space \space \Longleftrightarrow
\quad \  \pmb{\mathcal{D}^{-1}} \approx
\begin{bmatrix}
    \tau_{1} & 0 & 0 & \dots  & 0 \\
    0 &\tau_{2} & 0 & \dots  & 0 \\
    \vdots & \vdots & \vdots & \ddots & \vdots \\
    0 & 0 & 0 & \dots  & \tau_{N}
\end{bmatrix}.
\end{equation} However, the off-diagonal terms in the $\pmb{\mathcal{D}}$ matrix characterize the phonon-phonon mode scatterings, and they are usually not zero even within the approximation of the PG model. Similarly,  the cross-correlation  functions between two vibrational modes are usually not  zero even within the approximation of the PG model.

	The analytical solution of the probability function of an OU type vibration FPE also predicts the time-correlation functions of multiple vibrational modes. For example, based on the derivation in Appendix \ref{sec:Ab}, all the correlation  functions of odd-number vibrational modes are zero for an OU type lattice. There are three types of four-mode correlation functions:
\begin{equation} \label{eq:4cf_1}
\begin{split}
& \langle \delta n_\alpha(0) \delta n_\beta(0) \delta n_\mu(0) \delta n_\nu(t) \rangle _{eq}  = {(\Delta_{\alpha,eq} \Delta_{\beta,eq} \Delta_{\mu,eq} \Delta_{\nu,eq}) }^{\frac{1}{2}}  \\
&  \cdot \sum_{\lambda \lambda{'} \lambda{''} \lambda{'''}} (u_{\lambda,\alpha} u_{\lambda{'},\beta} u_{\lambda{''},\mu} u_{\lambda{'''},\nu}  ) \cdot \langle x_{\lambda}(0) \cdot x_{\lambda '}(0) \cdot x_{\lambda ''}(0) \cdot x_{\lambda '''}(t) \rangle _{eq} \\
= & {(\Delta_{\alpha,eq} \Delta_{\beta,eq} \Delta_{\mu,eq} \Delta_{\nu,eq}) }^{\frac{1}{2}} \cdot [\delta_{\alpha \mu} c_{\beta \nu}(t) +  \delta_{\beta \mu} c_{\alpha \nu}(t) + \delta_{\alpha \beta} c_{\mu \nu}(t)],\\
\end{split}
\end{equation}
	  
\begin{equation} \label{eq:4cf_2}
\begin{split}
& \langle \delta n_\alpha(0) \delta n_\beta(t) \delta n_\mu(t) \delta n_\nu(t) \rangle _{eq}  =  {(\Delta_{\alpha,eq} \Delta_{\beta,eq} \Delta_{\mu,eq} \Delta_{\nu,eq}) }^{\frac{1}{2}}\\
&  \cdot \sum_{\lambda \lambda{'} \lambda{''} \lambda{'''}} (u_{\lambda,\alpha} u_{\lambda{'},\beta} u_{\lambda{''},\mu} u_{\lambda{'''},\nu}  ) \cdot \langle x_{\lambda}(0) \cdot x_{\lambda '}(t) \cdot x_{\lambda ''}(t) \cdot x_{\lambda '''}(t) \rangle _{eq} \\
= &{(\Delta_{\alpha,eq} \Delta_{\beta,eq} \Delta_{\mu,eq} \Delta_{\nu,eq}) }^{\frac{1}{2}} \cdot [\delta_{\mu \nu} c_{\alpha \beta}(t) +  \delta_{\beta \nu} c_{\alpha \mu}(t) + \delta_{\beta \mu} c_{\alpha \nu}(t)],\\
\end{split}
\end{equation}

\begin{equation} \label{eq:4cf_3}
\begin{split}
& \langle \delta n_\alpha(0) \delta n_\beta(0) \delta n_\mu(t) \delta n_\nu(t) \rangle _{eq}  =  {(\Delta_{\alpha,eq} \Delta_{\beta,eq} \Delta_{\mu,eq} \Delta_{\nu,eq}) }^{\frac{1}{2}} \\
&  \cdot \sum_{\lambda \lambda{'} \lambda{''} \lambda{'''}} (u_{\lambda,\alpha} u_{\lambda{'},\beta} u_{\lambda{''},\mu} u_{\lambda{'''},\nu}  ) \cdot \langle x_{\lambda}(0) \cdot x_{\lambda '}(0) \cdot x_{\lambda ''}(t) \cdot x_{\lambda '''}(t) \rangle _{eq} \\
= & {(\Delta_{\alpha,eq} \Delta_{\beta,eq} \Delta_{\mu,eq} \Delta_{\nu,eq}) }^{\frac{1}{2}} \cdot [\delta_{\alpha \beta} \delta_{\mu \nu} + c_{\alpha \mu}(t) \cdot c_{\beta \nu}(t)  +   c_{\alpha \nu}(t) \cdot c_{\beta \mu}(t)], \\
\end{split}
\end{equation} with $c_{\alpha \beta}(t)$ being the normalized time-correlation function between $\alpha$-th mode and $\beta$-mode (Eqs. \ref{eq:normal_twomode}, \ref{eq:ou_mode_corr} and \ref{eq:integral_mode_corr}), and  the initial values of the four-mode time correlation functions derived as $\langle \delta n_\alpha(0) \delta n_\beta(0) \delta n_\mu(0) \delta n_\nu(0) \rangle _{eq}  = {(\Delta_{\alpha,eq} \Delta_{\beta,eq} \Delta_{\mu,eq} \Delta_{\nu,eq}) }^{\frac{1}{2}} \cdot [\delta_{\alpha \beta} \delta_{\mu \nu}  +  \delta_{\alpha \mu}\delta_{\beta \nu}  +  \delta_{\alpha \nu} \delta_{\beta \mu}]$.  

\section{\label{sec:kappa}Lattice Thermal Conductivity}

\subsection{\label{sec:gk}Green-Kubo Theory}
	The fluctuation-dissipation theorem provides a general statistical theory to connect the equilibrium fluctuation processes of a macroscopic quantity e.g.  the total heat current vector $\vec{J}=(J_x, J_y, J_z)$ in a solid and the related irreversible transport processes, such as heat conduction at non-equilibrium conditions.   Within the statistical linear response transport theory, the thermal conductivity tensor $\kappa_{IJ}$, with $I, J =x,y,z$ labeling the Cartesian axes, is expressed in the Green-Kubo formula in terms of the time integral of the current-current correlation functions \cite{green1954markoff, kubo1957statistical}:  
\begin{equation} \label{eq:gk}
\kappa_{IJ} = \dfrac{1}{k_B T^2 \Omega N_{cell}} \int_0^\infty dt \langle J_I(0) J_J(t) \rangle _{eq},
\end{equation} where $\Omega$ and $N_{cell}$ are respectively  volume of the unit-cell  and total number of cells in a super-cell model with the Born-von Karman periodic boundary.

	At the atomistic level, the heat current $\vec{J}$ is a function of atomic forces, displacements and momenta, and various approximations have been proposed and discussed \cite{hardy1963energy}. Assuming the heat current vector is also a function of phonon numbers of modes, i.e. $\vec{J} = \vec{J} ( \{ \Gamma \}) = \vec{J}(n_1, n_2, n_3, \cdots, n_N )$, we can use Eq. \ref{eq:correlation} to evaluate the current-current correlation functions. Under the condition of small thermal fluctuation, the Cartesian components of the heat current vector can be simplified as:
\begin{equation} \label{eq:J_linear}
J_I  \approx \sum_\alpha \dfrac{\partial J_I}{\partial \Delta n_\alpha} \Delta n_\alpha = \sum_i \Lambda_{I \alpha} \Delta n_\alpha. 
\end{equation} The seminal Peierls formula of the heat current of a phonon gas,  $\vec{J}  = \sum_\alpha \Delta n_\alpha \hbar \omega_\alpha \vec{v}_\alpha$,  is an approximation of this class, with $\Lambda_{I \alpha} = \hbar \omega_\alpha v_{\alpha I}$. When the higher order terms(also referred as the non-harmonic terms) in the $J$ formula are included as the corrections to the  linear terms formulated in Eq. \ref{eq:J_linear}, we can re-write the $J_I$ as $J_I =  \sum_i \Lambda_{I \alpha} \Delta n_\alpha + \delta J_I$.  Consequently,  the current-current correlation  functions can be expressed as: 
\begin{equation} \label{eq:J_corr}
\begin{split}
\langle J_I (0) J_J(t) \rangle _{eq} & = \sum_{\alpha \beta} \Lambda_{I \alpha} \Lambda_{J \beta} \langle \Delta n_\alpha(0) \Delta n_\beta(t) \rangle _{eq} \\
& + \sum_{\alpha} \Lambda_{I \alpha}  \langle \Delta n_\alpha(0) \delta J_{J}(t) \rangle _{eq}\\
& + \sum_{\alpha} \Lambda_{J \alpha}  \langle \delta J_{I}(0)  \Delta n_\alpha(t) \rangle _{eq}\\
& +  \langle \delta J_{I}(0)  \delta J_{J}(t) \rangle _{eq}.\\
\end{split} 
\end{equation} 

	Wherever the non-harmonic $\delta \vec{J}$ terms in the vibrational heat current in a lattice are not negligible, time-correlation functions of multiple modes, such as the four-mode correlation functions shown in Eq. \ref{eq:4cf_1}, \ref{eq:4cf_2}, \ref{eq:4cf_3}, are needed to evaluate the current-current correlation  function shown in Eq. \ref{eq:J_corr}.  At the condition that  the general linear approximation of Eq. \ref{eq:J_linear} is valid, the time integral of  $\langle J_I (0) J_J(t) \rangle _{eq}$ is approximated in terms of  time-integrals of normalized two-mode correlation functions $c_{\alpha \beta}(t)$:  
\begin{equation} \label{eq:J_linear_int}
\begin{split}
\int_0^\infty dt \langle J_I (0) J_J(t) \rangle _{eq} & \approx \sum_{\alpha \beta} \Lambda_{I \alpha} \Lambda_{J \beta}  \int_0^\infty dt \langle \Delta n_\alpha(0) \Delta n_\beta(t) \rangle _{eq} \\
& = \sum_{\alpha \beta} \Lambda_{I \alpha} \Lambda_{J \beta} (\Delta_{\alpha,eq} \cdot \Delta_{\beta,eq})^{1/2} \int_0^\infty dt c_{\alpha \beta}(t).
\end{split}
\end{equation}
	
	Based on the GK formula, we now express $\kappa_{Latt}$ in the form of: 
\begin{equation} \label{eq:gk_linear}
\kappa_{IJ}  =  \dfrac{1}{k_B T^2 \Omega N_{Cell}} \sum_{\alpha \beta} \Lambda_{I \alpha} \Lambda_{J \beta} (\Delta_{\alpha,eq} \cdot \Delta_{\beta,eq})^{1/2} \int_0^\infty dt c_{\alpha \beta}(t).
\end{equation}


	As shown in Eq. \ref{eq:ou_mode_corr}  of  Sec. \ref{sec:oua}, when a lattice vibration can be approximated as an  Ornstein-Uhlenbeck process, the lattice thermal conductivity is simply:
\begin{equation} \label{eq:kappa_fpe}
\begin{split}
\kappa_{IJ}  & = \dfrac{1}{k_B T^2 \Omega N_{Cell}} \sum_{\alpha \beta} \Lambda_{I \alpha} \Lambda_{J \beta}    (\Delta_{\alpha,eq} \cdot \Delta_{\beta,eq})^{1/2}  \sum_{\lambda = 1}^{N}    \int_0^\infty dt  e^{-\gamma_{\lambda}t} u_{\lambda,\alpha} u_{\lambda, \beta} \\
& =  \dfrac{1}{k_B T^2 \Omega N_{Cell}} \sum_{\alpha \beta} \Lambda_{I \alpha} \Lambda_{J \beta}    (\Delta_{\alpha,eq} \cdot \Delta_{\beta,eq})^{1/2}  \sum_{\lambda = 1}^{N}  (\gamma_{\lambda})^{-1} u_{\lambda,\alpha} u_{\lambda, \beta} \\
& =  \dfrac{1}{k_B T^2 \Omega N_{Cell}} \sum_{\alpha \beta} \Lambda_{I \alpha} \Lambda_{J \beta}    (\Delta_{\alpha,eq} \cdot \Delta_{\beta,eq})^{1/2}  (\pmb{\mathcal{D}})^{-1}_{\alpha \beta}. 
\end{split}
\end{equation}

\subsection{\label{sec:phbte} Phonon Boltzmann Transport Equation}
	As a kinetic transport theory, the phonon BTE theory is valid only within the PG approximation, i.e. at a thermal equilibrium,  each mode oscillates at a harmonic  frequency $\omega$ and the ensemble averaged number of phonons at this  mode follows the Bose-Einstein  distribution $n_{eq}(\omega) =\langle n \rangle_{eq} = \frac{1}{e^(\hbar \omega / k_{B}T) -1}$ and $\Delta_{eq}= \langle n^2 \rangle_{eq} - \langle n \rangle_{eq}^2= n_{eq} \cdot (n_{eq} + 1)$.  In addition, the phonon BTE theory applies only to a crystalline solid, where each vibrational mode of this translation-invariant periodic lattice corresponds to a  reciprocal-space  $\vec{k}$ vector and  a  group velocity $\vec{v}= \vec{\bigtriangledown}_k(\omega)$.
	
	When a constant temperature gradient $\vec{\bigtriangledown}_{r}T$ is imposed on the periodic lattice, the ensemble averaged phonon numbers,  $n_\alpha$ for $\alpha = 1, 2, 3,  \cdots, N$,  are no longer able to relax back to their original equilibrium values $n_{eq,\alpha}$ as a result of  thermal diffusion. Instead, each $n_\alpha$  approaches a  space-dependent value when a steady-state is reached:  

\begin{equation} \label{eq:phbte1}
\left(\dfrac{\partial n_\alpha}{dt}\right) = -\left(\dfrac{dn_\alpha}{dt}\right)_{diffusion} - \left(\dfrac{dn_\alpha}{dt}\right)_{scattering} =0, 
\end{equation} where  the diffusion term at the  $\vec{\bigtriangledown}_{r}T \rightarrow 0$ limit is approximated as:

\begin{equation} \label{eq:phbte_d}
\left(\dfrac{dn_\alpha}{dt} \right)_{diffusion}  =   - \vec{v}_\alpha  \cdot \vec{\bigtriangledown}_{r} n_\alpha  \simeq  -  \dfrac{\hbar \omega_\alpha}{k_B T^2} n_{\alpha,eq}(n_{\alpha,eq} + 1) \vec{v}_\alpha  \cdot \vec{\bigtriangledown}_{r} T.
\end{equation}

	A common approximation for the scattering terms in the phonon BTE (Eq. \ref{eq:phbte1}) is the so-called linearized  approximation:
\begin{equation} \label{eq:phbte_s}
 \left(\dfrac{dn_\alpha}{dt} \right)_{scattering} \simeq - \sum_{\beta=1}^{N} \sqrt{\frac{n_{\alpha,eq} \cdot (n_{\alpha,eq} + 1)}{n_{\beta,eq} \cdot (n_{\beta,eq} + 1)}} \cdot \mathcal{L}_{\alpha \beta}  \cdot (n_\beta - n_{\beta, eq}), 
\end{equation} where $\pmb{\mathcal{L}}$ is referred as the linear phonon scattering matrix. 

	By using the results  of Eqs. \ref{eq:phbte_d} and \ref{eq:phbte_s} and the definition of  $\phi_\alpha \equiv \dfrac{n_\alpha  - n_{\alpha,eq}  } {\sqrt{n_{\alpha,eq} \cdot (n_{\alpha,eq} + 1) }}$,  the steady-state phonon Boltzmann equation (Eq.  \ref{eq:phbte1}) can be re-written as a set of linear equations for $\phi_\alpha$ with $\alpha =1, 2, 3,  \cdots, N$:
\begin{equation} \label{eq:phbte2}
\sum_{\beta=1}^{N}  \mathcal{L}_{\alpha \beta}  \cdot \phi_\beta  =   -  \dfrac{\hbar \omega_\alpha}{k_B T^2} \sqrt{n_{\alpha,eq}(n_{\alpha,eq} + 1)} \vec{v}_\alpha  \cdot \vec{\bigtriangledown}_{r} T.
\end{equation} 

	Similar to what we have derived in Sec. \ref{sec:oua}, we can solve the set of linear equations using the eigenvectors and the eigenvalues of the matrix $\pmb{L}$:
\begin{equation} \label{eq:phbte_solution}
\begin{split}
\phi_\alpha  & = - \sum_{\lambda = 1}^{N} {\gamma_\lambda}^{-1} u_{\lambda,\alpha} \sum_{\beta = 1}^{N} \dfrac{\hbar \omega_\beta}{k_B T^2} \sqrt{n_{\beta,eq}(n_{\beta,eq} + 1)} (\vec{v}_\beta  \cdot \vec{\bigtriangledown}_{r} T) u_{\lambda,\beta} \\
& =   - \sum_{\beta =1}^{N}   (\sum_{\lambda = 1}^{N} {\gamma_\lambda}^{-1} u_{\lambda,\alpha} u_{\lambda,\beta} )  \cdot \dfrac{\hbar \omega_\beta}{k_B T^2} \sqrt{n_{\beta,eq}(n_{\beta,eq} + 1)} (\vec{v}_\beta  \cdot \vec{\bigtriangledown}_{r} T) \\
& =   - \sum_{\beta =1}^{N}  (\pmb{\mathcal{L}})^{-1}_{\alpha \beta}  \cdot \dfrac{\hbar \omega_\beta}{k_B T^2} \sqrt{n_{\beta,eq}(n_{\beta,eq} + 1)} (\vec{v}_\beta  \cdot \vec{\bigtriangledown}_{r} T), 
\end{split} 
\end{equation}  where  $\gamma_{\lambda}$ and $\vec{u_{\lambda}}$ are the $\lambda$-th eigenvalue and eigenvector of the matrix $\pmb{\mathcal{L}}$, and $(\pmb{\mathcal{L}})^{-1}$ represents  the inverse matrix of $\pmb{\mathcal{L}}$.

	Based on the Peierls formula for the heat current of a phonon gas, the lattice thermal conductivity predicted by the linearized phonon BTE theory can be expressed as:
\begin{equation} \label{eq:kappa_bte}
\begin{split}
 \kappa_{I J} & =      \dfrac{1}{\Omega N_{Cell}} \sum_{\alpha \beta}    (\pmb{L})^{-1}_{\alpha \beta}  \cdot  \dfrac{\sqrt{n_{\alpha,eq}(n_{\alpha,eq} + 1)}  \sqrt{n_{\beta,eq}(n_{\beta,eq} + 1)} \hbar \omega_\alpha \hbar \omega_\beta}{k_B T^2}   {v}_{\alpha I} \cdot {v}_{\beta J}\\
  & =  \dfrac{1}{\Omega N_{Cell}} \sum_{\alpha \beta} {(c_\alpha c_\beta)}^{1/2}  \cdot {v}_{\alpha I} \cdot {v}_{\beta J} \cdot (\pmb{L})^{-1}_{\alpha \beta},  \\
\end{split}
\end{equation} where $c = k_{B} \cdot (\frac{\hbar \omega}{k_{B}T})^{2} \cdot n_{\alpha,eq} \cdot (n_{\alpha,eq} +1)$ is the single mode heat capacity.

	To compare $\kappa_{Latt}$ predicted by the phonon BTE (Eq. \ref{eq:kappa_bte}) and the one by the OU type vibration FPE (Eq. \ref{eq:kappa_fpe}), we first note that in the limit of weak phonon scattering of the PG model, the variance of the phonon number fluctuation of a mode $\Delta_{\alpha,eq}$ has already been shown to converge to  the value of $n_{\alpha,eq} \cdot (n_{\alpha,eq}+1)$, and the Peierls formula of heat current is valid. Furthermore, with the interpretation of phonon occupation number $n_{\alpha}$ in the phonon BTE as the time-dependent expectation value of the phonon number during the thermal relaxation process, we conclude that the normalized drift/diffusion matrix $\pmb{\mathcal{D}}$ in an OU type vibration FPE (Eq. \ref{eq:ou_A}) is identical to the linear phonon scattering matrix  $\pmb{\mathcal{L}}$, i.e. $\pmb{\mathcal{D}} \rightarrow \pmb{\mathcal{L}}$, at the weak phonon scattering limit of the PG approximation. Consequently, $\kappa_{Latt}$ predicted by the vibration FPE (Eq. \ref{eq:kappa_fpe}) converges to that predicted by the conventional phonon BTE (Eq. \ref{eq:kappa_bte}). The so-called single mode relaxation  approximation  (SMRA) or relaxation time approximation  (RTA) of a kinetic transport model corresponds to the cases where the phonon scattering matrix $\pmb{\mathcal{L}}$ (or  the drift/diffusion matrix $(\pmb{\mathcal{D}}$) can be treated as a positively defined diagonal matrix (Eq. \ref{eq:diagonal_L}).  

\subsection{\label{sec:discussions}Discussions}
	 
	 A comparison chart is shown in Table \ref{tab:tab1} to highlight commonality and distinction between the atomistic MD simulation method and the vibration FPE.  The  MD simulation approach has an absolute advantage in simulating the atomistic scale  lattice heat currents at moderate and high temperature,  and it applies consistently to disordered solids, very anharmonic solids, as well as  fluids. However,  MD simulations  only provide a semi-quantitative  description of the fluctuation properties of  individual vibrational modes based on the damped oscillator model. Firstly,  corrections to the quantized lattice vibration have to be considered at low temperature because of the classical nature of MD simulations. Secondly, the mode lifetimes extracted from the numerical solutions of MD trajectories over long  simulation periods reflect only partial information on the fluctuation  and relaxation processes in lattice dynamics.  Because of the assumption that all the cross-mode correlation functions  between two different vibrational modes are zero, the damped oscillator approximation is equivalent to the single mode relaxation approximation or relaxation time approximation  in  kinetic transport theories.  The predicted $\kappa_{Latt}$  from these approximate kinetic theories are known to be  noticeably underestimated comparing to those derived from the full solutions of the phonon BTE theory at  low temperature \cite{ward2009ab,LI20141747} or in low dimension materials \cite{PhysRevX.6.041013}.

\begin{table}[]
\caption{Comparison chart for the molecular dynamics method and the vibration Fokker-Planck equation method. }
\label{tab:tab1}
\resizebox{\textwidth}{!}{\begin{tabular}{l|l|l}
\hline
& \multicolumn{1}{c|}{\begin{large}\textbf{Molecular Dynamics Simulation}\end{large}}  & \multicolumn{1}{c}{\begin{large}\textbf{Fokker-Plank Equation}\end{large}}  \\ \hline

\begin{tabular}[c]{@{}c@{}}\textbf{Thermal}\\ \textbf{Conductivity}\\ $\kappa$\end{tabular} & \multicolumn{2}{c}{\begin{tabular}[c]{@{}c@{}}\begin{large} \textit{Green-Kubo Formula}:\end{large} \begin{large}\(\displaystyle\kappa_{IJ} = \dfrac{1}{\Omega k_B T^2 } \int_0^\infty \langle J_\alpha(0) J_\beta(t) \rangle_{eq} dt \) \textit{with J(t) as the fluctuating heat current} \end{large} \\ \begin{large} \textit{Required Input}: the time-correlation function of heat current $\langle J_\alpha(0) J_\beta(t) \rangle _{eq}$.\end{large}\end{tabular}}                                                                                                                                                            \\ \hline

\begin{tabular}[c]{@{}c@{}}\textbf{Heat}\\ \textbf{Current}\\ $J$\end{tabular} &
\begin{tabular}[c]{@{}l@{}} \(\displaystyle J =\dfrac{1}{2} \sum_i \sum_{j > i} \textbf{F}_{ij} \cdot (\textbf{v} _i+ \textbf{v}_j)(\textbf{r}_i - \textbf{r}_i) \) \\ \(\textbf{r}_i, \textbf{v}_i \sim \textit{atomic position/velocity} \) \\ \(\textbf{F}_{ij} \sim \textit{force between two atoms} \) \end{tabular} &
\begin{tabular}[c]{@{}l@{}} \( \displaystyle J(\{n_1, n_2, \cdots, n_{N}) = \textbf{J}_{0} + \Delta \textbf{J}, n \sim \textit{phonon numbers}  \) \\ \( \textit{Harmonic heat current: } J_{0} = \sum_{\alpha} n_{\alpha} \hbar \omega_{\alpha} \textbf{v}_{g \alpha}, (\omega,v_g) \sim \textit{frequency/group velocity} \)  \\ \( \textit{Anharmonic heat current: } \Delta \textbf{J} \rightarrow \textit{beyond the Perierls approximation, not yet formulated} \) \end{tabular}
\\ \hline 

\textbf{Dynamics} &
\begin{tabular}[c]{@{}l@{}}\( \textit{Classical Newton's $2^{nd}$ Law} \)  \\
\( \displaystyle \textbf{F}_i = -  \nabla _{\textbf{r}_i} V(\textbf{r}_1, \cdots ,\textbf{r}_n) \)\\ 
\(V \sim \textit{many-body inter-atomic potential} \) 
\end{tabular}    &
\begin{tabular}[c]{@{}l@{}} \(  \textit{Quantum, Semi-Classical, and Classical} \)  \\ \( \displaystyle A_{\alpha}(\Gamma)  \equiv \lim_{\delta t \rightarrow 0} \frac{1}{\delta t} \int_{0}^{\delta t} d\Gamma^{'} \delta n_\alpha(\Gamma, \Gamma^{'})P(\Gamma^{'},\delta t \vert \Gamma) \) \\ \(\displaystyle B_{\alpha\beta}(\Gamma)  \equiv \lim_{\delta t \rightarrow 0} \frac{1}{\delta t} \int_{0}^{\delta t} d\Gamma^{'} \delta n_\alpha(\Gamma, \Gamma^{'}) \delta n_\beta(\Gamma,\Gamma^{'})P(\Gamma^{'},\delta t \vert \Gamma) \) \end{tabular}
\\ \hline

\begin{tabular}[c]{@{}c@{}}\textbf{Phonon}\\ \textbf{Gas}\\ \textbf{Limit}\end{tabular}          &                                                                                                                                                                                                                                                                                                                                                             
\begin{tabular}[c]{@{}l@{}}  \textit{Damped Oscillators and Mode Lifetimes} \\ \( \displaystyle V= V_{harmonic} + \Delta V_{anh}, \Delta V_{anh} \gg V_{harmonic} \)\\  Only numerical solution for $\textbf{r}_i(t)$, $\textbf{v}_i(t)$ \\  \textit{Normal mode analysis}:\\
$\bullet$ Long simulation time is necessary \\ $\bullet$ Self-correlation functions $c_{\alpha \alpha}(t)$ often \\ \quad assumed to be $e^{-t/ \tau_{\alpha}} $\\ $\bullet$ Cross correlations $c_{\alpha \beta} (t)$ often assumed \\ \quad to be zero  \\ $\bullet$ Three/four-mode correlation functions \\ \quad rarely discussed\end{tabular}
&            
\begin{tabular}[c]{@{}l@{}} \( \textit{OU Approximation of drift/diffusuion coefficients} \Longleftrightarrow \textit{Phonon Gas Model} \) \\ \( A_\alpha(\Gamma) = - \sum_{\beta} \mathcal{D}_{\alpha \beta} \sqrt{{\Delta_{\alpha,eq}}/{\Delta_{\beta,eq}})} \delta n_\beta, B_{\alpha \beta}(\Gamma) = 2\sqrt{\Delta_{\alpha,eq} \cdot \Delta_{\beta,eq})}  \mathcal{D}_{\alpha \beta} \)\\ 
\textit{Analytical Solution}: \( P(\tilde{\Gamma}, t |{\tilde{\Gamma}^0}) = \prod_{\lambda  =1} ^N \dfrac{1}{\sqrt{2\pi\Delta_\lambda (t)}} e^{-\dfrac{[x_\lambda - \overline{x_{\lambda}}(t)]^2}{2\Delta_\lambda (t)}} \)\\ \textit{Analytical formulas of multiple-mode time-correlation function}:
\\$\bullet$ Two-mode: \( c_{\alpha \beta} = \langle \delta n_{\alpha}(0) \cdot \delta n_{\beta} (t) \rangle_{eq} / \sqrt{\langle \delta n_{\alpha}^2 \rangle_{eq} \cdot \langle \delta n_{\beta}^2 \rangle_{eq}} \) \\ \quad \quad \quad \quad \quad \quad \quad  \ \  \( \displaystyle =\sum_{\lambda = 1}^N e^{-\gamma_{\lambda} t} u_{\lambda,\alpha} u_{\lambda, \beta} \Rightarrow \int_0^\infty d\gamma \chi_{\alpha \beta}(\gamma) e^{-\gamma t} \) with $N \rightarrow \infty$
\\ $\bullet$ Three-mode: \( \langle \delta n_\alpha(0) \delta n_\beta(0) \delta n_\mu(t)\rangle _{eq} = 0\)
\\ $\bullet$ Four-mode: \( \langle \delta n_\alpha(0) \delta n_\beta(0) \delta n_\mu(0) \delta n_\nu(t) \rangle _{eq} =\)
\\ \quad \([\delta_{\alpha \mu} c_{\beta \nu}(t) +  \delta_{\beta \mu} c_{\alpha \nu}(t) + \delta_{\alpha \beta} c_{\mu \nu}(t)] \sqrt{\langle \delta n_{\alpha }^2 \rangle_{eq} \langle \delta n_{\beta }^2 \rangle_{eq}  \langle \delta n_{\mu }^2 \rangle_{eq}  \langle \delta n_{\nu }^2 \rangle_{eq} }  \)  \end{tabular}            \\ \hline
\end{tabular}}
\end{table}
	 
	  In contrast, the vibration FPE approach complements the conventional MD simulation approach for conditions in which the interactions among vibrational modes are moderate, and it can be adopted  to delineate the breakdown conditions of the PG model in MD simulations. Based on vibration FPE, we propose that  the PG model applies when the OU approximation of the drift and diffusion coefficients  (Eqs. \ref{eq:ou_A} and \ref{eq:ou_B}) is valid.  By considering  the normalized drift/diffusion $\pmb{\mathcal{D}}$  matrix in an OU type vibration FPE equivalent to the scattering $\pmb{\mathcal{L}}$ matrix in a phonon BTE, we have proved for the first time that the $\kappa_{Latt}$ derived from the linear response transport theory converges to that from the kinetic transport theory within the PG approximation. 
	  
	  When the interactions among vibrational modes are perturbatively small, the normalized drift/diffusion $\pmb{\mathcal{D}}$  matrix can be derived by using quantum perturbation theories for lattice vibration at low temperature. As temperature elevates to the semi-classical and classical regime, we can implement numerical algorithms to directly compute normalized drift/diffusion coefficients  with first-principles MD simulations.  As these coefficients are defined in the short time limit,  high-performance parallel computer  platforms can be utilized to distribute the computational loads of such simulations in parallel.  When the temperature dependence of the drift/diffusion coefficients are extracted  and tested with the OU approximation, we are able to not only quantitatively determine the temperature condition in which the PG model breaks down, but also identify the individual vibrational modes that lead to the breakdown.

\section{\label{sec:conculusions}Conclusions}
	In summary, we have developed a vibration Fokker-Planck equation theory to describe stochastic lattice dynamics in solids. Instead of  simulating the atomic trajectories using the molecular dynamics methods, this statistical theory  characterizes the fluctuation and relaxation processes  in terms of a time-dependent,  multiple-mode probability function, evolving from a thermally sampled single micro-state at $t=0$ (Eq. \ref{eq:initial_prob}) to the equilibrium distribution over all the accessible micro-states as $t \rightarrow \infty$ (Eq. \ref{eq:eq_prob}). The  dynamical properties that govern the stochastic processes at  atomistic scale are coarse-grained into two  sets of parameters of a vibration FPE, the drift  $A$ and diffusion $B$ coefficients of vibrational modes (Eqs. \ref{eq:phFPE} and \ref{eq:fpe_AB}).  At the limit of weak mode-mode interactions, these coefficients can be derived  with quantum perturbation theories, such as the Fermi's golden rule (Eq. \ref{eq:fpe_AB_perturbation} and \ref{eq:goldenrule}).  Beyond the perturbation approximation, these coefficients can  be directly computed by using MD methods over short simulation time periods (i.e. $\delta t \approx 0$). Thus, the intensive computational loads of sampling a large amount of initial micro-states of a vibrating lattice can be distributed in a computer platform with massive parallel algorithms.   
	
	Our time-dependent probability theory presents a new paradigm to compute correlation functions among vibrational modes (Eqs. \ref{eq:normal_twomode} and \ref{eq:dt_normal_twomode}). The advantage of this statistical approach is clearly demonstrated at the Ornstein-Uhlenbeck condition (Sec. \ref{sec:oua}), in which the vibration FPE has an analytical solution (Eqs. \ref{eq:ou_prob} - \ref{eq:transform}) and the  correlation functions among multiple modes (Eqs. \ref{eq:ou_mode_corr}, \ref{eq:4cf_1} - \ref{eq:4cf_3}) can be derived in terms of eigenvalues and eigenvectors of the normalized drift/diffusion matrix $\pmb{\mathcal{D}}$ (Eqs. \ref{eq:ou_A} and \ref{eq:ou_B}). By equating the $\pmb{\mathcal{D}}$ matrix in an OU type vibration FPE with the conventional phonon scattering matrix $\pmb{\mathcal{L}}$ (Eq. \ref{eq:phbte_s}) in a phonon BTE, we have presented the first rigorous mathematical proof to equalize $\kappa_{Latt}$ results from both the Green-Kubo theory (Eq. \ref{eq:kappa_fpe}) and the BTE theory (Eq. \ref{eq:kappa_bte}) with the  Peierls harmonic heat current formula (Eq. \ref{eq:J_linear}).
	
	Although both the vibration FPE theory and the phonon BTE theory predict identical $\kappa_{Latt}$ results  within the PG model, the vibration FPE provide additional theoretical insights on the heat conduction mechanism at microscopic level. Firstly, the vibration FPE theory quantatitively defines the contributions to the overall $\kappa_{Latt}$ from  both the  self-correlaltion functions of individual modes and the cross-correlation functions between two different modes (Eq. \ref{eq:gk_linear}). Secondly,  the vibartion FPE further predicts all the multiple-mode correlation functions, which can be analyzed in future to account effects of anharmonic correction terms in heat flux \cite{hardy1963energy,PhysRevB.82.224305}. Finally, when perturbation theories become insufficient to  evaluate the phonon scattering matrix $\pmb{\mathcal{L}}$ of a phonon BTE, the full set of matrix elements of $\pmb{\mathcal{L}}$, instead of merely effective phonon lifetimes,  can  be computed as the normalized drift/diffusion coefficients of an OU type FPE by using the MD simulations over short time periods.

	To study the mechanisms of lattice heat conduction beyond the PG model, it is critical to establish a quantatitive criterion that delineates  the breakdown conditions. The theoetical analysis presented in this paper indicates that the OU condition of stochastic lattice dynamics (Eqs. \ref{eq:ou_A} and \ref{eq:ou_B}) might serve as such breakdown criterion. We are currently implmenting MD methods to compute  the temperaure-depedent drift/diffusion coefficients up to the melting temperature of a laltice. Various numerical methods, such as adiabatic  elimination of variables method, matrix continued-fraction method, or variational methods, will be examined to solve the vibration FPE byond the OU approximation\cite{risken1996fokker}. It is promising  that this  vibration FPE  presents a new theoretical framework to accurately and effectively predict the stochastic vibrational processes and the thermal transport properties of solids within and beyond the PG model.

\begin{acknowledgments}
This work is financially supported by the National Sciences Foundation with the grant EAR-1346961. JD thanks the Thomas and Jean Walter Professor endowment from Department of Physics, Auburn University. YZ acknowledges the Alabama Commission on Higher Education for the support of Graduate Research Student Program Fellowships (Rounds 10-12). The authors also thank J.D. Perez, D. Crawford, D.A. Drabold,  O.F. Sankey, and B. Xu for discussions.
\end{acknowledgments}

\appendix
\section{Expectation values and statistical variances of numbers of phonons} \label{sec:Aa}
	This appendix provides some derivation details on some formulas about the expectation values and statistical variances of the phonon numbers  shown in Sec. \ref{sec:sd}.
	
	We first define the time-dependent expectation values of the following three quantities using the ensemble average approach shown in Sec. \ref{sec:relaxation}: 

\begin{equation} \label{eq:A1}
\overline{n}_\alpha (t|\Gamma^0)  \equiv \langle n_\alpha \rangle = \int d\Gamma n_\alpha P(\Gamma; t),
\end{equation}

\begin{equation} \label{eq:A2}
\overline{n_\alpha n_\beta}(t|\Gamma^0)  \equiv \langle n_\alpha n_\beta \rangle = \int d\Gamma n_\alpha n_\beta  P(\Gamma; t),
\end{equation}

\begin{equation} \label{eq:A3}
\overline{\Delta}_{\alpha \beta}(t|\Gamma^0) = \langle n_\alpha n_\beta \rangle - \langle n_\alpha  \rangle \langle n_\beta \rangle .
\end{equation}

	Using the vibration FPE shown in  Eq. \ref{eq:phFPE},  we then prove that the  first-order $t$-derivatives of these three quantities in Eqs. \ref{eq:A1} to \ref{eq:A2} have the following forms: 
\begin{equation} \label{eq:A4}
\begin{split}
\dfrac{d \overline{n}_\alpha}{d t} & = \int d \Gamma n_\alpha \dfrac{\partial}{\partial t}P(\Gamma,t|\Gamma^0) \\
& = - \sum_i \int d \Gamma n_\alpha \dfrac{\partial}{\partial n_i}[A_i(\Gamma) \cdot P(\Gamma, t|\Gamma^0)] \\
&  + \dfrac{1}{2} \sum_{ij} \int d\Gamma        n_\alpha \dfrac{\partial^2}{\partial n_i \partial n_j}[B_ij(\Gamma) \cdot P(\Gamma,t|\Gamma^0)].
\end{split}
\end{equation}

	For  $\alpha \neq i$, we have $\int d\Gamma n_\alpha \dfrac{\partial}{\partial n_i} [A_i(\Gamma) P(\Gamma, t|\Gamma^0)] = 0$. Similarly, $\int d\Gamma n_\alpha \frac{\partial}{\partial n_i} \frac{\partial}{\partial n_j} [B_{ij}(\Gamma) P(\Gamma, t|\Gamma^0)] = 0$ for $\alpha \neq i$ or $\beta \neq j$.  As a result, Eq. \ref{eq:A4} is now simplified as:  

\begin{equation} \label{eq:A5}
\dfrac{d \overline{n}_\alpha (t|\Gamma^0)}{d t}  =  -  \int d \Gamma n_\alpha \dfrac{\partial}{\partial n_\alpha}(A_\alpha\cdot P) + \dfrac{1}{2}  \int d\Gamma n_\alpha \dfrac{\partial^2}{\partial n_\alpha \partial n_\alpha}(B_{\alpha\alpha} \cdot P), \\
\end{equation}  with 
$\int d\Gamma n_\alpha \dfrac{\partial}{\partial n_\alpha} (A_\alpha \cdot P) = \int d \Gamma \dfrac{\partial}{\partial n_\alpha} (n_\alpha \cdot A_\alpha \cdot P) - \int d (\Gamma A_\alpha \cdot P)  =- \int d\Gamma A_\alpha(\Gamma) \cdot P(\Gamma, t|\Gamma^0) = - \overline{A_\alpha}(t|\Gamma^0)$, and 
$\int d\Gamma n_\alpha \dfrac{\partial}{\partial n_\alpha} \dfrac{\partial}{\partial n_\alpha} (B_{\alpha\alpha} \cdot P)  = \int d\Gamma  \frac{\partial}{\partial n_\alpha} [n_\alpha \dfrac{\partial}{\partial n_\alpha} (B_{\alpha\alpha} P)] - \int d \Gamma \dfrac{\partial}{\partial n_\alpha}(B_{\alpha\alpha} P) = 0$. We now get:

\begin{equation} \label{eq:A6}
\dfrac{d \overline{\alpha}(t|\Gamma^0)}{d t} =   \overline{A_\alpha}(t|\Gamma^0).
\end{equation}

	Similarly, we can show that 
\begin{equation} \label{eq:A7}
\dfrac{d \overline{n_\alpha n_\beta}(t|\Gamma^0)}{d t} =   \overline{n_{\alpha} A_\beta}(t|\Gamma^0) + 	\overline{n_{\beta} A_\alpha}(t|\Gamma^0) + \overline{B_{\alpha \beta}}(t|\Gamma^0). 
\end{equation} Consequently, we also have:

\begin{equation} \label{eq:A8}
\begin{split}
\dfrac{d \overline{\Delta_{\alpha\beta}}(t|\Gamma^0)}{d t} & =  \dfrac{d \overline{n_\alpha n_\beta}(t|\Gamma^0)}{d t}  - \overline{n_\alpha}(t|\Gamma^0) \cdot \dfrac{d \overline{n_\beta}(t|\Gamma^0)}{d t}  - \overline{n_\beta}(t|\Gamma^0) \cdot \dfrac{d \overline{n_\alpha}(t|\Gamma^0)}{d t}  \\
 & = \overline{B_{\alpha \beta}}(t|\Gamma^0) + [\overline{n_{\alpha} A_\beta}(t|\Gamma^0) - \overline{n_{\alpha}}(t|\Gamma^0) \cdot \overline{A_{\beta}}(t|\Gamma^0) ] + [ \overline{n_{\beta} A_\alpha}(t|\Gamma^0) - \overline{n_{\beta}}(t|\Gamma^0) \cdot \overline{A_{\alpha}}(t|\Gamma^0)].
\end{split}
\end{equation} 

\section{Analytical solutions of the FPE for an OU process} \label{sec:Ab}
	In this appendix we verify the analytical solutions of an OU type FPE (Eq. \ref{eq:ou_fpe}) shown in Sec. \ref{sec:oua}. For a probability function of one stochastic variable $x$ with zero-mean and unit-variance, the corresponding OU type FPE can is given as:
\begin{equation}\label{eq:B1}
\frac{\partial P(x,t)}{\partial t} = \gamma \cdot ( 1 +  x \cdot \dfrac{\partial}{\partial x} +  \dfrac{\partial^2}{\partial^2 x} ) P(x,t). 
\end{equation} 

	With $\frac{d}{dt}\overline{x} = - \gamma  \overline{x}$,  $\frac{d}{dt} \overline{x^2} = 2 \gamma (1- \overline{x^2} $), and the initial values $\overline{x}$ and $\overline{x^2}$ being $x_0$ and $(x_0)^2$ respectively,  we have:
\begin{equation} \label{eq:B2}
\begin{split}
\overline{x} & = x_{0} e^{-\gamma t},  \\
\Delta & = \overline{x^2} - {\overline{x}}^2 = 1 - e^{-2\gamma t}.
\end{split}
\end{equation}

	We skip the details of derivation and only show that the analytical solution of  Eq. \ref{eq:B1} is given as:
\begin{equation}\label{eq:B3}
P(x,t) =  \dfrac{1}{\sqrt{2 \pi \Delta(t)}} e^{- \dfrac{(x - \overline{x}(t))^2}{2\Delta(t)}}. 
\end{equation}

	Using the analytical solution in Eq. \ref{eq:B3}, we can show that $\frac{\partial P(x,t)}{\partial t} $ on the left hand side of Eq. \ref{eq:B1} is:
\begin{equation} \label{eq:B4}
\begin{split}
\frac{\partial P(x,t)}{\partial t}  & = \frac{\partial [\dfrac{1}{\sqrt{2 \pi \Delta(t)}}]}{\partial t}  e^{- \dfrac{(x - \overline{x}(t))^2}{2\Delta(t)}}  + \dfrac{1}{\sqrt{2 \pi \Delta(t)}} \frac{\partial  [e^{- \dfrac{(x - \overline{x}(t))^2}{2\Delta(t)}}] }{\partial t} \\
& = \gamma [1 - \Delta^{-1}(t)]P(x,t)  + \gamma [\dfrac{(x-\overline{x}(t))^2}{\Delta^2(t)} - \dfrac{x \cdot (x - \overline{x}(t)}{\Delta(t)} ] P(x,t) \\
& = \gamma [1 - \Delta^{-1}(t) \cdot (1  - x \overline{x}(t) + x^2) + \Delta^{-2}(t) \cdot (x-\overline{x}(t))^2 ] P(x,t).
\end{split}
\end{equation}

	From the analytical solution in Eq. \ref{eq:B2}, we also have $ \frac{\partial P(x,t)}{\partial x} = -\Delta^{-1}(t) \cdot (x-\overline{x}(t)) \cdot P(x,t)$  and  $ \frac{\partial^2 P(x,t)}{\partial x^2} = -\Delta^{-1}(t)  + \Delta^{-2}(t) \cdot (x-\overline{x}(t))^2 \cdot P(x,t)$. These results give us the right hand side of Eq. \ref{eq:B1} in the form of:
\begin{equation} \label{eq:B5}
\begin{split}
& \quad \ \gamma \cdot [ 1 +  x \cdot \dfrac{\partial}{\partial x} +  \dfrac{\partial^2}{\partial^2 x} ] P(x,t)\\
& = \gamma \cdot [1 - x \cdot \Delta^{-1}(t) \cdot (x-\overline{x}(t))  -\Delta^{-1}(t)  + \Delta^{-2}(t) \cdot (x-\overline{x}(t))^2 ] \cdot P(x,t)\\
& = \gamma \cdot [1 - \Delta^{-1}(t) \cdot (1 - x \cdot \overline{x}(t) + x^2)  + \Delta^{-2}(t) \cdot (x-\overline{x}(t))^2 ] \cdot P(x,t). 
\end{split}
\end{equation}	

	With both Eq. \ref{eq:B4} and Eq. \ref{eq:B5}, we now verify that the probability function in Eq. \ref{eq:B3} is indeed the analytical solution of Eq. \ref{eq:B1}.
	
	For the case $N > 1$, the $N$-dimensional probability function of an OU type FPE (Eq. \ref{eq:ou_fpe}) can be expressed in a separable form $P(\tilde{\Gamma}, t) = \prod_{\lambda  =1} ^N f_\lambda(x_\lambda,t)$, and one $N$-variable FPE (Eq. \ref{eq:ou_fpe}) is converted into $N$ sets of partial differential equations:
\begin{equation}\label{eq:B6}
\frac{\partial f_\lambda (x,t)}{\partial t} = \gamma_\lambda \cdot ( 1 +  x_\lambda \cdot \dfrac{\partial}{\partial x_\lambda} +  \dfrac{\partial^2}{\partial^2 x_\lambda} ) f_\lambda (x,t),
\end{equation} where $\lambda = 1, 2, 3, \cdots N$. Similar to the solution shown in Eq. \ref{eq:B3}, we have the $N$ sets of solutions of $f_\lambda (x,t) =  {(2 \pi \Delta_\lambda)}^{-1/2} \cdot e^{- \dfrac{(x_\lambda - \overline{x_\lambda})^2}{2\Delta_\lambda}}$, with $\overline{x_\lambda}  = x_{\lambda,0} e^{-\gamma_\lambda t}$ and $\Delta_\lambda  = 1 - e^{-2\gamma_\lambda t}$. Plugging these results in the separable multiple-variable formula, we now can verify that the analytical solution of Eq. \ref{eq:ou_fpe} is indeed the probability function shown in Eq. \ref{eq:ou_prob}.

	The analytical solutions of the probability function for an OU type FPE allows us to directly derive the correlation functions among these state variables with zero-means and unit-variances. For example, the time correlation functions between any two stochastic variables can be shown as the following familiar forms:   
\begin{equation} \label{eq:B7}
\begin{split}
& \langle x_{\lambda}(t^{'}) \cdot x_{\lambda '}(t^{'} + t) \rangle _{eq}   =  \langle x_{\lambda}(0) \cdot x_{\lambda '}(t) \rangle _{eq} \\
& = \langle x_{\lambda} \cdot x_{\lambda '} \rangle _{eq} e^{-\gamma_{\lambda '} t}  =  \delta_{\lambda,\lambda^{'}} e^{-\gamma_\lambda t}. \\
\end{split}
\end{equation}

	Meanwhile, we can prove that all three-variable correlation functions for a multiple variable  OU process zero:
\begin{equation} \label{eq:B8}
\begin{split}
\langle x_{\lambda}(0) \cdot x_{\lambda '}(0) \cdot x_{\lambda ''}(t)  \rangle _{eq}
& = \langle x_{\lambda} \cdot x_{\lambda '}  \cdot x_{\lambda ''} \rangle _{eq} e^{-\gamma_{\lambda ''} t} \\
& =  0 \cdot e^{-\gamma_{\lambda ''} t} = 0, \\
\end{split}
\end{equation}

\begin{equation} \label{eq:B9}
\begin{split}
\langle x_{\lambda}(t) \cdot x_{\lambda '}(0) \cdot x_{\lambda ''}(t)  \rangle _{eq} 
& = \langle x_{\lambda} \cdot x_{\lambda '}  \cdot x_{\lambda ''} \rangle _{eq} e^{-2\gamma_{\lambda ''} t} + \langle x_{\lambda} \rangle _{eq} \cdot (1 - e^{-2\gamma_{\lambda ''} t}) \cdot \delta_{\lambda{'} \lambda{''}} \\
& =  0 \cdot  e^{-2\gamma_{\lambda ''} t} + 0 \cdot (1- e^{-2\gamma_{\lambda ''} t}) \cdot \delta_{\lambda{'} \lambda{''}} = 0 . \\
\end{split}
\end{equation}

	We can further generalize that the correlation functions of  odd-number variables, such as three-variable, five-variable, etc, are all zero. Meanwhile, the correlation functions of  even-number variables are not always zero. For example, for four-variable correlation functions, we have the following free formula:
\begin{equation} \label{eq:B10}
\begin{split}
\langle x_{\lambda}(0) \cdot x_{\lambda '}(0) \cdot x_{\lambda ''}(0) \cdot x_{\lambda '''}(t) \rangle _{eq} & = \langle x_{\lambda} \cdot x_{\lambda '}  \cdot x_{\lambda ''} \cdot x_{\lambda '''} \rangle _{eq} e^{-\gamma_{\lambda '''} t} \\
=  & (\delta_{\lambda \lambda{'''}} \cdot \delta_{\lambda {'} \lambda{''}} +  \delta_{\lambda{'} \lambda{'''}} \cdot \delta_{\lambda \lambda{''}}  + \delta_{\lambda{''} \lambda{'''}} \cdot \delta_{\lambda \lambda{'}} ) e^{-\gamma_{\lambda {'''}} t},  \\
\end{split}
\end{equation}		

\begin{equation} \label{eq:B11}
\begin{split}
& \langle x_{\lambda}(0) \cdot x_{\lambda '}(t) \cdot x_{\lambda ''}(t) \cdot x_{\lambda '''}(t) \rangle _{eq} \\
=  & \langle x_{\lambda} \cdot x_{\lambda '}  \cdot x_{\lambda ''} \cdot x_{\lambda '''} \rangle _{eq} e^{-(\gamma_{\lambda '} + \gamma_{\lambda ''}+ \gamma_{\lambda '''} ) t}  +  \langle x_{\lambda} x_{\lambda{'}} \rangle _{eq} \cdot e^{-\gamma_{\lambda{'}} t} \cdot (1 - e^{-2\gamma_{\lambda{''}} t} ) \cdot \delta_{\lambda{''} \lambda{'''}}  +   \\
& \langle x_{\lambda} x_{\lambda{''}} \rangle _{eq} \cdot e^{-\gamma_{\lambda{''}} t} \cdot (1 - e^{-2\gamma_{\lambda{'}} t} ) \cdot \delta_{\lambda{'} \lambda{'''}}  +   \langle x_{\lambda} x_{\lambda{'''}} \rangle _{eq} \cdot e^{-\gamma_{\lambda{'''}} t} \cdot (1 - e^{-2\gamma_{\lambda{'}} t} ) \cdot \delta_{\lambda{'} \lambda{''}}   \\
= &  (\delta_{\lambda \lambda{'}} \cdot \delta_{\lambda {''} \lambda{'''}} +  \delta_{\lambda \lambda{''}} \cdot \delta_{\lambda{'} \lambda{'''}}  + \delta_{\lambda \lambda{'''}} \cdot \delta_{\lambda{'} \lambda{''}} )  \cdot e^{-(\gamma_{\lambda{'}} + \gamma_{\lambda {''}} + \gamma_{\lambda {'''}}) t}   + \\
&  e^{-\gamma_{\lambda} t} \cdot [ \delta_{\lambda \lambda{'}} \cdot \delta_{\lambda {''} \lambda{'''}} \cdot (1- e^{ -2 \gamma_{\lambda{''}} t})  +  \delta_{\lambda \lambda{''}} \cdot \delta_{\lambda {'} \lambda{'''}} \cdot (1- e^{ -2 \gamma_{\lambda{'}} t}) +  \delta_{\lambda \lambda{'''}} \cdot \delta_{\lambda {'} \lambda{''}} \cdot (1- e^{ -2 \gamma_{\lambda{'}} t})  ]  \\
=  & (\delta_{\lambda \lambda{'}} \cdot \delta_{\lambda {''} \lambda{'''}} +  \delta_{\lambda \lambda{''}} \cdot \delta_{\lambda{'} \lambda{'''}}  + \delta_{\lambda \lambda{'''}} \cdot \delta_{\lambda{'} \lambda{''}} ) e^{-\gamma_{\lambda} t},  \\
%
\end{split}
\end{equation}

\begin{equation} \label{eq:B12}
\begin{split}
& \langle x_{\lambda}(0) \cdot x_{\lambda '}(0) \cdot x_{\lambda ''}(t) \cdot x_{\lambda '''}(t) \rangle _{eq}  \\
= & \langle x_{\lambda} \cdot x_{\lambda '}  \cdot x_{\lambda ''} \cdot x_{\lambda '''} \rangle _{eq} e^{-(\gamma_{\lambda ''}+ \gamma_{\lambda '''} ) t}  + \langle x_{\lambda} x_{\lambda{'}} \rangle _{eq} \cdot (1 -  e^{-2\gamma_{\lambda ''} t}) \cdot \delta _{\lambda{''} \lambda{'''}}\\
=  &  (\delta_{\lambda \lambda{''}} \cdot \delta_{\lambda {'} \lambda{'''}} +  \delta_{\lambda \lambda{'''}} \cdot \delta_{\lambda{'} \lambda{''}}  + \delta_{\lambda \lambda{'}} \cdot \delta_{\lambda{''} \lambda{'''}} ) e^{-(\gamma_{\lambda {''}} + \gamma_{\lambda {'''}}) t} + \delta_{\lambda \lambda{'}} \cdot \delta_{\lambda {''} \lambda{'''}} \cdot (1- e^{-2 \gamma_{\lambda{''}} t }) \\
 = &  \delta_{\lambda \lambda{'}} \cdot \delta_{\lambda {''} \lambda{'''}}  +  (\delta_{\lambda \lambda{''}} \cdot \delta_{\lambda {'} \lambda{'''}} +  \delta_{\lambda \lambda{'''}} \cdot \delta_{\lambda{'} \lambda{''}}  ) e^{-(\gamma_{\lambda} + \gamma_{\lambda {'}}) t}.
\end{split}
\end{equation}

\bibliographystyle{apsrev4-1}
\bibliography{vibFPE_ref}

\end{document}